\documentstyle[aas2pp4]{article}

\begin{document}

\title{HUBBLE SPACE TELESCOPE IMAGING OF the CFRS and LDSS REDSHIFT SURVEYS--- III. Field elliptical galaxies
at $0.2 < z < 1.0$\footnotemark[1]$^,$\footnotemark[2]}

\author{David Schade$^a$, S.J. Lilly$^b$,D. Crampton$^a$, R. S. Ellis$^c$,O. Le F\`evre$^{d,e}$
Francois Hammer$^e$, Jarle Brinchmann$^c$, R. Abraham$^c$, Matthew Colless$^f$, K. Glazebrook$^g$, L. Tresse$^c$,
Tom Broadhurst$^i$}  
\affil{$^a$National Research Council Canada, HIA/DAO, 5071 West Saanich Road, Victoria V8X 4M6, Canada\\
$^b$Dept. of Astronomy, University of Toronto, 60 St. George St., Toronto M5S 3H8, Canada\\
$^c$Institute of Astronomy, Madingley Road, CB3 OHA Cambridge, England\\
$^d$ Laboratoire d'Astronomie Spatiale, Traverse du Siphon, B.P.S. 13376 Marseille, Cedex 12, France\\
$^e$Observatoire de Paris, Section de Meudon, DAEC, 92195 Meudon Principal Cedex, France\\
$^f$Research School of Astronomy and Astrophysics,Australian National University, Canberra, ACT 0200, Australia\\
$^g$Anglo-Australian Observatory, Siding Spring, Coonabarabran, NSW 2357, Australia\\
$^i$Astronomy Department, University of California, Berkeley, CA 94720, USA\\
}
%
%
%

\footnotetext[1]{Based on observations with the NASA/ESA {\it Hubble Space 
Telescope}
obtained at the Space Telescope Science Institute, which is operated by the
Association of Universities for Research in Astronomy Inc., under NASA
contract NAS 5-26555. }
\footnotetext[2]{Based in part on data obtained through the facilities
of the Canadian Astronomy Data Centre. }

\begin{abstract}

Two-dimensional surface photometry has been performed on a magnitude-limited sample 
of 46 field galaxies that are classified as ellipticals based on two-dimensional fitting of
their luminosity profiles using Hubble Space Telescope imaging. These galaxies are
described well by a deVaucouleurs $R^{1/4}$ profile. The
sample was selected from the combined Canada-France and
LDSS redshift surveys and spans the redshift range 0.20$<z<$1.00.
This analysis reveals several clear evolutionary trends.
First, the relationship between galaxy half-light radius and
luminosity evolves with
redshift such that a galaxy of a given size is more luminous by
$\Delta M_B=-0.97 \pm 0.14$ mag at $z=0.92$ relative to the
local cluster elliptical relation. Second, the mean rest-frame color shifts blueward
with redshift by $\Delta (U-V) =-0.68 \pm 0.11$ at  $z=0.92$ relative to the
same relation in the Coma cluster. These shifts in color and luminosity 
of field elliptical galaxies are similar to those measured for cluster ellipticals.
Approximately 1/3 of these elliptical
galaxies (independent of redshift) exhibit [OII] 3727 emission lines with 
equivalent widths $> 15$  angstroms indicating ongoing star formation. Therefore,
field elliptical galaxies are not composed entirely of very old stellar populations. 
Estimated star-formation rates together with stellar population evolutionary models
imply that $\le 5\%$ of the stellar mass in the elliptical galaxy population has
been formed since $z=1$. We find some evidence that
the dispersion in color among field ellipticals at $z\sim 0.55$ may be 
larger than that seen among samples of cluster ellipticals and S0 galaxies at similar 
redshift. We see no evidence for a decline in the space density of early-type
galaxies with look-back time. The $<V/V_{max}>$ statistics and a comparison
with local luminosity functions are both consistent with the 
view that the population of massive early-type galaxies was largely in place by $z\sim 1$.
This implies that merging 
is not required since that time to produce the present-day space  density
of elliptical galaxies. However, the statistics are poor: a larger sample is
required to produce a decisive result.

\end{abstract}


\section{INTRODUCTION}

 The simplest models of galaxy evolution have massive elliptical galaxies forming at
high redshift in a rapid collapse and monolithic burst of star formation. Such scenarios
have been described by Eggen, Lynden-Bell, \& Sandage (1962) and Larson (1975).
Thereafter, elliptical galaxy stellar populations
age passively, with no further star formation. Hierarchical clustering
models (White \& Rees 1978, Kauffman, White, \& Guiderdoni 1993) require that 
the most massive objects form at late times via the merging of
smaller subunits. In this scenario, massive ellipticals would be assembled recently
and would be absent from high-redshift galaxy surveys. At redshift $z=1$ 
the space density would be lower a factor of 2-3 (Kauffmann  1996,
Baugh, Cole, \& Frenk 1996) compared to $z=0$ 
for $\Omega_\circ=1$ (with a less dramatic
decrease expected for smaller $\Omega_\circ$).
These two views of the
assembly of massive ellipticals (early monolithic
formation and recent merging) are diametrically opposed to one another.

 The monolithic collapse/passive evolution
scenario will be referred to as the "orthodox" model because of
the rigidity of its prohibition of any further star formation following the
initial burst at formation. This prohibition renders the color and 
luminosity evolution that
accompany the aging of the galaxy more easily predictable by theoretical means.
 The late-epoch merging or hierarchical clustering model will be referred to as the 
"secular" model. It is also useful to define an intermediate or "reform" model
which has massive ellipticals assembling most of their stellar mass early
(e.g., $z > 3$) but where some fraction of the stars form later ($z < 2$).
The reform model
is useful both because it represents a plausible physical scenario and because
it helps to characterise the sensitivity of different types of observational
tests of elliptical galaxy formation and evolution. 

 In summary, the orthodox and reform models both require the assembly of the majority of
the stellar mass of elliptical galaxies at high redshift whereas recent merging plays
a central role in elliptical galaxy formation in the secular model.

A large fraction of the work on elliptical galaxies has been concentrated on
the cluster environment where early-type galaxies are the dominant population. That work
may provide useful insights into the origin and evolution of the field
population but those studies might also be misleading if environment plays a major role
in galaxy formation and evolution.

\subsection{Evidence in favor of the orthodox view}

In many respects, cluster ellipticals appear to
form a remarkably homogeneous population. The existence of a
well-defined color-luminosity relation for early-type galaxies
suggests a degree of homogeneity in that population since color
depends on age and metallicity. Bower, Lucey, \& Ellis (1992)
show that the dispersion in the color-luminosity
relation in Coma is small enough to require
either a high formation redshift ($z\ge  3$) or a high degree of synchronicity in
galaxy formation times. 

 The existence of a low-scatter fundamental plane for early-type galaxies in clusters
(Djorgovski \& Davis 1987, Dressler et al. 1987) based on size, 
surface brightness, and velocity dispersion further implies
a degree of regularity in the mass-to-light ratios, and therefore the stellar
populations, among the early-type population. Jorgensen, Franx, 
\& Kjaergaard (1996) find no dependence of the fundamental plane
coefficients or scatter on environment (e.g., cluster richness or velocity
dispersion) a conclusion supported by Pahre, de Carvalho, \& Djorgovski (1998).
 Guzman \& Lucey (1993) disagree, suggesting that ellipticals in lower
density environments may have slightly younger stellar populations than
those residing in regions of higher density.
Forbes, Ponman, \& Brown (1998) analyse a sample 
of (mostly field) ellipticals and find a correlation between the time since
the last major episode of star formation and fundamental plane residuals.
This argues that {\em field} ellipticals span a wide range in age. 

More recently, measurements of the early-type galaxy population have
been pushed out to higher redshift.
The well-defined color-luminosity relation in clusters at $z=0.55$
(Ellis et al. 1997) shows that the dispersion among early-type galaxies
is small even at that epoch.
However, the debate is still active regarding exactly how
strong a constraint on elliptical galaxy formation is presented by
the colour-magnitude relation (Kauffman \& Charlot 1998, Bower, Kodama,
\& Terlevich 1998).

Studies of the cluster elliptical fundamental
plane (van Dokkum \& Franx 1996, Kelson et al. 1997, van Dokkum et al. 1998) 
and the weakening of the Mgb absorption
as a function of redshift (Bender, Ziegler, \& Bruzual 1996) support 
simple passively-evolving (orthodox) models.
Purely photometric techniques (Pahre, Djorgovski, \& de Carvalho 1996,
Barrientos, Schade, \& L\'{o}pez-Cruz 1996, Schade et al. 1996) have
been use to compile larger samples of ellipticals covering a wide range in redshift.
Schade, Barrientos \& L\'{o}pez-Cruz (1997) show
that the surface brightness of cluster elliptical galaxies to $z\sim 1$ evolves
substantially ($\Delta \mu(B) \sim -1$ mag) in a manner broadly consistent with
early-forming and passively-evolving models (e.g., Bruzual \& Charlot 1993).

Color evolution of cluster ellipticals
(Dressler and Gunn 1990, Aragon-Salamanca et al. 1993, Rakos and Schombert 1995,
 Oke, Gunn, \& Hoessel 1996, Standford, Eisenhardt, \& Dickinson 1997) is
found to be broadly consistent with passive evolution of an old stellar population
with little recent star formation.

Lilly et al. (1995) examined the distributions of $<V/V_{max}>$ for
a photometrically-defined sample of early-type galaxies (i.e., the subset of red galaxies)
in the Canada-France Redshift Survey (CFRS) and found that
this population is distributed in space in a manner consistent with
a constant space density to $z\sim 1$.

The evidence cited above suggests that "orthodox" 
models of elliptical galaxy formation provide at least broad agreement 
with the observed properties of the cluster elliptical population over a 
range ($z=0$ to $z\sim 1$) in redshift. In the field, the situation is
less well-defined. Sandage \& Visvanathan (1978) show that
the $z=0$ field and cluster color-luminosity relations are very similar
and have comparably small scatter ($\sigma \simeq 0.10$ mag) implying
an evolutionary history that does not depend strongly on environment.
Schade et al. (1996) use ground-based imaging to compare cluster and field ellipticals 
to $z=0.55$ and find roughly similar amounts of luminosity evolution in the two
environments. On the other hand the work by Forbes, Ponman, \& Brown (1998)
and Guzman \& Lucey (1993) suggests evolutionary histories that depend
on environment, a result supported by Mobasher \& James (1996).

\subsection{Evidence against the orthodox view}

 Observational tests of purely passive evolution come in two flavors. In
the first, it is only the orthodox view that can be rejected. That is, the result
cannot be used to discriminate between the reform or secular (merging) model. 
The second flavor
constitutes a direct test of the merging model and can provide a rejection of
both the orthodox and reform points of view.

 Examples of the first flavor of test are studies that search for high-redshift galaxies
that have the colors predicted by population synthesis models for passively-evolving 
stellar systems. 
Kauffman, Charlot, \& White (1996) claim evidence for strong evolution of the
E/S0 population over $0.2 < z < 1.0$ from a $V/V_{max}$  analysis of the 
Canada-France Redshift Survey 
(CFRS; Lilly et al. 1995). Selection of a sample of early-type 
(i.e., E or S0) galaxies was attempted
on the basis of optical colors predicted by evolving population synthesis models.
Note that Kauffman, Charlot, \& White compare the sample selected in this way
to the predictions of semi-analytic galaxy formation models for
the number density of objects with the {\it morphological} properties of
ellipticals. Their claim of strong evolution requires either that early-type
galaxies are absent at high redshift (which would support the secular model) or
that they are present but have colors that are blue enough to avoid detection
by the particular color selection that was applied to compile the sample. The latter case
would support the reform model since the blue color is presumably produced
by ongoing star formation.

Totani \& Yoshii (1998) have repeated an analysis of the CFRS
sample similar to the analysis by Kauffman, Charlot, \& White (1996) 
but do not support the conclusions
of the earlier work. In particular, Totani \& Yoshii use different evolutionary models
to select early-type galaxies and they restrict the sample redshift range to
$0.2 < z < 0.8$. The analysis yields results consistent with passive evolution
models without merging.

There are other studies searching for high-redshift galaxies with
passively-evolving colors.
Zepf (1997) concludes from a deep search for objects with very red 
optical-infrared colors that typical
ellipticals must have had significant star formation more recently than
$z=5$. Barger et al. (1998) use infrared observations to estimate that roughly 50\% of the 
local space density of elliptical galaxies was in place with passively-evolving
colors at $1.3 < z < 2.2$. Menanteau et al. (1998) conclude that there is a deficit
of a factor of 3 or more in the observed number of red ellipticals 
compared to monolithic collapse models. Abraham et al. (1998) find limited evidence
for a diversity of star-formation histories among a sample of 12 bright,
morphologically-selected spheroids in the Hubble Deep Field but do not
attempt to determine space density variations with redshift.

 As has been noted, the observational tests described in the preceding
several paragraphs are in principle capable of rejecting
only the orthodox model even if the experiments are perfectly executed. In other words,
they may fail to detect the expected number of galaxies at high redshift because 
either a) the galaxies are absent (because they form via merging at an epoch later
than the epoch of observation) or b) the galaxies are present but bluer than the threshold color.
These experiments rely on theoretical models to predict the expected colors. Therefore,
actual galaxies may be bluer than the theoretical threshold because they have undergone
recent star formation (and thus violate the tenets of the orthodox model) or because
the theoretical colors for passively-evolving populations are in error. 
For example, Jimenez et al.
(1998) present an alternative model for the evolution of a stellar population
where more than 90\% of the stellar mass of
ellipticals is formed in the first gigayear of their existence yet these
galaxies are substantially bluer than standard population synthesis models at
intermediate redshifts. This study illustrates that uncertainties in modelling
old stellar populations may be a significant contributor to uncertainties
in the conclusions drawn from searches for very red galaxies at intermediate
redshifts.

\subsection{Evidence in favor of the merging scenario}
 
 Theories based on hierarchical cluster models of structure formation
(e.g. White \& Rees 1978)
naturally predict that massive galaxies will be the last galaxies to
form by merging. Numerical simulations of disk-disk mergers 
(White \& Negroponte 1973, Barnes \& Hernquist 1996) showing that
the remnants have a number of properties similar to elliptical galaxies 
give this scenario further plausibility. On the observational side, studies of close
pairs of galaxies (Zepf \& Koo 1989, Carlberg, Pritchet, \& Infante 1994, Patton
et al. 1997) indicate that the galaxy interaction or merging rate rises
steeply with redshift ($\sim(1+z)^3$). This result is supported by a study
of the pair fractions and merging rate in the CFRS sample (Le F\`evre et al.
1999).

Schweizer \& Seitzer (1992) studied the relation between morphological
features and colors among local early-type galaxies.
They find evidence that the colors of field and group E and S0
galaxies are correlated with the presence of fine morphological
structure believed to be produced by merging events. They argue
that the systematic shift toward bluer colors with increased
structure is likely produced by systematic variations in the
mean age of the stellar population.
The elapsed times
since the last merger events are estimated to be in the 3-10 Gyr range.

 The elliptical galaxy population contains many examples 
of galaxy cores that are kinematically 
decoupled from the main body of the galaxy (Bender 1992 and
references therein). Although this is clearly evidence against monolithic
formation of the entire stellar population in ellipticals, it is not
clear whether it requires major mergers or whether more minor interactions
might be responsible (see, e.g., Hau \& Thompson 1994).

 At high redshift there are tests 
 that bypass color information (and thus
bypass the uncertainties in the population synthesis models) by 
selecting early-type galaxies directly on the basis of morphology.

Counts of galaxies as a function of morphological type (Driver et al. 1998)
in the the Hubble Deep Field (HDF) provide an indication at roughly the
$2\sigma$ level of a
deficit of elliptical galaxies at $z > 1$ relative to purely passively-evolving
models. Franceschini et al. (1998) use a K-band selected sample together with
morphological information to define a sample of 35 elliptical and S0 galaxies
in the HDF with spectroscopic or photometric redshifts. These authors suggest that
the episodes of star formation that produce field ellipticals are spread over
a wide range in redshift ($1 < z < 4$) rather than concentrated in a single
burst at any epoch (a similar result is found by Kodama, Bower, \& Bell 1998). 
They also note a deficit of early-type galaxies at $z > 1.3$.

The HDF samples are small and originate in a single small field and are so are susceptible to
the effects of clustering along the line of sight.\footnotemark[3] 
\footnotetext[3]{As a check on this problem a subsample matching the selection criteria of the
CFRS sample ($I_{AB} \le 22.5$) can be extracted from the Franceschini et al. sample.
 Depending on the morphology classification chosen (including or excluding S0s) and
the assumed redshift of completeness for the CFRS ($z=0.8$ or $z=1.0$) the ratio
of Franceschini to CFRS early-type galaxies per unit area is in the range 2.2 to
2.9 indicating a significant surplus of  early-type galaxies with $I_{AB} \le 22.5$ in the HDF 
(relative to the CFRS which is averaged over many lines of sight) up to $z=1$.}Clearly
multiple lines of sight offer distinct advantages.

\subsection{Evidence against the merging scenario}

In an attempt to confirm the results of
Schweizer \& Seitzer (1992), Silva \& Bothun (1998a) used near-infrared imaging to
search for an intermediate-age (1-3 Gyr) stellar
population among a sample of ellipticals with morphological signatures of 
mergers (largely overlapping the Schweizer \& Seitzer sample). They find
that these galaxies cannot be distinguished from those ellipticals without 
signs of interaction and that a small fraction ($< 15\%$ at most) of the
stellar mass can be attributed to an intermediate-age population. The test
is based on the search for an intermediate-age asymptotic giant-branch
population. The search for evidence of mergers is extended (Silva \& Bothun
1998b) into the central regions of the galaxies where gas-rich 
disk-disk mergers would be expected to produce a concentration of younger
stars but no compelling evidence for such a population is found. These
authors argue that any merging that has occurred has not been accompanied
by a strong burst of star formation, either distributed globally or concentrated
in the central regions of the galaxies.

A measurement of the space density of large ellipticals
as a function of look-back time could produce a direct resolution to the
issue of merging as an important process in forming elliptical galaxies. 
Density evolution can be 
detected using either the $V/V_{max}$ statistic (which tests the
hypothesis that a population is distributed uniformly in distance) 
or by direct comparison of estimated space densities of large ellipticals
at high and low redshift. If merging is important for producing
early-type galaxies then there must be fewer
large elliptical galaxies at high redshift and a correspondingly 
low value of the $<V/V_{max}>$ statistic. 

 In a key piece of work 
Im et al. (1996) construct the luminosity function (LF) for a sample
of 376 E/S0 galaxies selected by morphology and luminosity-profile analysis
but {\em independent of color}. They use HST imaging and photometric redshifts based
on a training set of their own and CFRS redshifts and detect evolution
in the characteristic luminosity of the early-type LF.  They find 
a value of $<V/V_{max}>=0.58\pm 0.01$ for their entire sample.
 Note that the high value of $<V/V_{max}>$ implies an excess probability
of finding ellipticals at high redshift. Presumably this excess is produced by the
positive luminosity evolution rather than by an increase in space density.
A correction for that evolution would reduce $<V/V_{max}>$.
 The value of $<V/V_{max}>$ and the form of the luminosity function are
both consistent with a constant space density  of elliptical galaxies
to $z\sim1$.
 The great strength of this study lies in the large number of galaxies used
which provides a firm statistical footing but the weakness lies in the
small numbers of spectroscopic redshifts used as a training set for the redshifts 
based on $V-I$ color, apparent magnitude, and half-light radius.

 In summary, there are strong theoretical reasons to believe that recent merging
may be an important process governing the assembly of massive galaxies.
Much of the evidence that very red (i.e. old and passively-evolving) galaxies
are deficient in space density at high redshift are based upon small samples so that the
statistical uncertainty is large. In addition there is uncertainty in
the theoretical models used to define the color threshold for the sample selection.
 In any case, studies based on color selection cannot discriminate between the
secular (merging) and reform models so they do not provide direct support for
the merging scenario. Comparisons of morphologically selected samples (number
counts or luminosity functions) disagree with one another. However, the Im et al. (1996) is
a strong piece of evidence that the space density of large ellipticals has
changed little since $z=1$.

\subsection{The present study}

 The definition of an elliptical galaxy used throughout this paper is
based on two-dimensional fitting procedures and is independent of
color. Only those galaxies
that are well-fit by an $R^{1/4}$ profile (de Vaucouleurs 1948) 
with no significant evidence of an additional disk component are
used. Those with asymmetry index greater than 0.10 (Schade et al. 1995; roughly the
fraction of the total flux contained in the asymmetric component of the galaxy
within 5 kpc of the galaxy center) have been excluded.

This is the third of a series of four
papers based on an amalgamation of redshifts and HST imaging
for galaxies in the Canada-France Redshift Survey (Lilly et al. 1995a
and references therein)
and LDSS Redshift Survey (Ellis et al. 1996).
The series forms a comprehensive study of the evolution
of galaxy populations to $z=1$ using a complete sample
of 341 galaxies with HST imaging. 
Brinchmann et al. (1998) (hereafter referred to as Paper I) discuss the morphological
properties of the sample as a whole and Lilly et al. (1998) 
(hereafter referred to as Paper II)
discuss the quantitative morphologies
of the late-type population. Le F\`evre et al. (1998) (Paper IV) deals
with the merging history of luminous galaxies.

 The present paper discusses the galaxy sample and the
image processing procedures (\S2), results for elliptical
galaxies are presented in
\S 3 and are discussed in \S 4. 
Throughout this paper (unless otherwise noted)
we have adopted $H_\circ=50h_{50}$ km sec$^{-1}$ Mpc$^{-1}$ and
$q_\circ=0.5$.

\section{DATA AND PROCEDURE}

\subsection{CFRS/LDSS Imaging }

 Details of the sample selection and imaging data are given in
Paper I and are summarised here. 
 HST imaging was obtained for a sample of 341 
galaxies from the
combined Canada-France Redshift Survey with
a limiting magnitude of $I_{AB}=22.5$ and the  
LDSS Redshift Survey (limit $b_j = 24$, Ellis et al. 1996) datasets. Typical
exposures were 6300 seconds in the F814W filter. A subset of the
fields also had F450W images of 6300 seconds duration. Archival
images in F814W and F606W (typically 4500 seconds) from the
region where the CFRS overlaps the ``Groth strip'' were also used.
 These overlaps provided multiple images for 38 galaxies.
 The typical signal-to-noise ratio (defined using the total flux
within a 3 arcsecond diameter aperture)
of the HST imaging is 100 at $I_{AB}=22$
and this is sufficient to allow visual classification or calculation
of concentration index (Paper I) or to allow two-dimensional surface
photometry. 

\subsection{Two-dimensional surface photometry}

 Given the modest signal-to-noise ratio of the high-redshift galaxy
observations, it is appropriate to adopt simple models to characterise
the luminosity profile of these galaxies. We adopt models
that have been shown to 
adequately represent the luminosity profiles of most 
galaxies in the nearby universe (e.g., de Vaucouleurs 1948,
Freeman 1970, Kormendy 1977, Kent 1985, Kodaira, Watanabe, \& Okamura 1986,
van der Kruit 1987, de Jong 1996, Courteau, de Jong, \& Broeils 1996).
 Two-dimensional surface photometry was performed on the HST
images via the fitting of two-component parametric models:
an exponential
disk (Freeman 1970) and $R^{1/4}$ bulge (deVaucouleurs 1948)
(see also de Jong 1996).
 These models were convolved with an empirically-defined point-spread-function
(PSF) and integrated over each pixel of the HST image. Free parameters
were size (scale length or half-light radius), axial ratio, and
position angle for each component, and fractional bulge luminosity
$B/T$. Together with the galaxy center (assumed to be coincident for
the two components) the fits thus require 9 parameters. Fitting was
done using a modified Levenberg-Marquardt algorithm (Press et al. 1992).
The technique is simple and robust and provides
structural and surface brightness information in addition to a morphological
classification. The problems of fitting two-component luminosity profiles
has been addressed by Simien \& de Vaucouleurs (1986) and Schombert \&
Bothun (1987).

 Real galaxies show a large variety of morphological features (bars, dust lanes,
spiral structure, companions) and are often asymmetric whereas the
 models adopted here are idealised and symmetric. 
In order to deal with these features
and also to allow the computation of an asymmetry index, each
galaxy was ``symmetrized'' by the following procedure (see also 
Elmegreen, B., Elmegreen, D., \& Montenegro 1992 and Schade
et al. 1995). First, the image of the galaxy was rotated by $180^\circ$ and subtracted from
itself. The resulting ``asymmetric'' image (with both positive and
negative asymmetric pixel values) was clipped so that only positive asymmetries
more significant than $2\sigma$ remained. This provides 
an estimate of the asymmetric component of the galaxy image and
the difference formed by subtracting this from the original image
yields the ``symmetric'' image. This symmetric image was subjected to
the fitting procedure and the ratio of the asymmetric to symmetric
flux is taken as an index of the relative importance of the asymmetric
component of the galaxy. From a practical standpoint, the symmetric
image is free of companions and other problems and results in a much
cleaner and reliable fitting process.

 Sky background was measured from large areas of the chips and was not
a parameter of the fit. Simulations showed that fitting the
sky as a
free parameter resulted in an increased scatter in the
fitted galaxy parameters but did not otherwise effect the results.
 With the sky held fixed, the models were
normalised at each iteration to yield a minimum $\chi^2$ given the current structural
parameters. 

 The fitting process results in measurements for each of the bulge and disk components
individually. This allows independent analyses of the evolution of the disk
components separately from the galaxy as a whole, or the bulge components
in isolation. Perhaps most importantly, the classifications 
(based solely on fractional bulge luminosity $B/T$) are
independent of color and spectroscopic properties so that no
implicit assumptions are made about the stellar populations in
individual galaxies. This approach allows a check of whether spectral energy
distribution and morphology are correlated at high redshift and
whether that correlation evolves.

 A crucial feature of the two-dimensional modelling as a means to
measure structural parameters and surface brightnesses is the fact
that we have applied these same techniques (Schade, Barrientos \& 
L\'{o}pez-Cruz 1997) to nearby elliptical
galaxies (albeit in clusters rather than the field) that were observed
with similar physical resolution and at similar signal-to-noise ratios.
This greatly reduces concerns that some of our trends are due to
systematic differences between the analyses of low and high redshift
samples.

\subsection{Visual examination of the fits}

 The evaluation of the success of the fitting process and the validity of
the model fits was done by a combination of statistics 
and visual examination of the galaxy, the models,
and, most importantly, the residuals. This procedure has subjective elements and needs a 
clear explanation so that the issue of possible bias can be addressed. 

The main function
of the visual inspection is to decide whether a single-component or
two-component (bulge-plus-disk) model is the best representation of the
galaxy under consideration. 
The smallest reduced $\chi^2$ is not always
accepted as the best fit.
Two-component models were rejected in favor 
of single component models if
 1) the fitted
bulge component was larger and of lower surface brightness than the disk
component (this is an explicit bias that we have imposed on
the procedure and was the most common type of
failure of the fitting process: it occurred in 3\% of the
galaxies), or 2) the second component was effectively
being fitted to low-level residual asymmetries (that escaped the
symmetrizing process) in the galaxy rather than a legitimate
bulge or disk component (this overlaps with reason 1). 

 The visual examination relies mainly on the residuals. If an examination
of the residuals after the bulge fit has been subtracted shows no evidence
for a disk component then the pure bulge fit will be accepted. This is true
even if, as happens in some cases, the bulge+disk model fit has a slightly 
smaller value of $\chi^2$. This subjective element of the procedure was 
evaluated by multiple examinations by one of us (DJS) of the fits 
separated by substantial periods 
of time. These repeated judgements showed that about 10\% of early-type
galaxies drifted from E to S0 or vice versa depending on whether a
disk component was considered a plausible. There are 65 early-type (E or S0) galaxies
so that the elliptical sample of 46 galaxies may have contamination of about
15\%. We adopt this as an estimate
of the reliability of our early-type classifications although we will
later examine the effect of 25\% contamination of the E sample which we
consider a limiting case.

\subsection{Internal errors from duplicate observations}

 There exist 38 duplicate galaxy observations. These are independent
observations (most being widely separated in time) which  
were reduced and analysed independently so that
 they give good estimates of the internal errors of 
our fitting procedure including potential sources of error such as
centroiding, PSF variations, sky subtraction, and the effect of the
subjective visual examination procedure. 

 The results of duplicate fits are shown in Figures 1 and 2. Figure
1 shows $B/T$, total $I$ magnitude, and the index of asymmetry. 
Filled symbols represent galaxies with good residuals (symmetric
or ``normal galaxies'') and open symbols indicate asymmetric galaxies
(those with asymmetry index as defined in Schade et al. 1995 of $R > 0.10$). 
It is important to note
that classifications are considered valid only for symmetric
galaxies and discussions of evolution of ``normal'' galaxies
in this paper pertains only to those galaxies; asymmetric objects
are excluded. The dispersions
are indicated on the figures for the 
{\em symmetric} galaxies only. The dispersions
for the symmetric (and all) galaxies are 0.04 (0.14) for $B/T$ with 70\% of the objects
having duplicate observations within $\pm 0.10$. In total magnitude, the
dispersion is 0.05 (0.11) magnitudes. The dispersion in the asymmetry index is 0.06. 
Since the index results in a binary decision (symmetric or asymmetric) it is 
appropriate to look at the errors in classification. A cutoff of in 
asymmetry index of 0.10 is adopted 
here and this results in 5/38 (13\%) disagreement between the two classifications,
whereas adopting a cutoff of 0.14 in asymmetry index (as adopted by Schade et al. 1995)
results in 1/38 (3\%) differences in classification. In summary, the repeat
observations show that the fitting procedure produces consistent measures
of $B/T$, total magnitude, and asymmetry index, with errors in the
range 5-15\%. 

 Figure 2 shows the multiple-fit results for structural parameters. The dispersions
for the symmetric (and asymmetric) disk scale length measurements are 0.04 (0.05)
arcseconds, or about 10\% for a typical disk, the surface brightness errors
are 0.20 (0.46) mag for disks, and the total disk magnitude errors are 0.10
(0.14) mag. The corresponding figures for the bulge (elliptical) measurements
are 0.04 (0.04) arcseconds in size, 0.10 (0.14) mag in surface brightness, and
0.04 (0.04) mag in total magnitude.

 The duplicate observations are a robust test of the entire analysis process
and include all sources of random errors. The size (scale length or half-light
radius) and magnitude errors are $\sim 10\%$ as inferred from these comparisons.
 The surface brightness measurement errors are $\sim 10-20\%$. 

\subsection{Magnitudes and  K-corrections}

 The HST $I_{814}$ magnitudes from the surface photometry 
were compared to the ground-based CFRS $I-band$ isophotal imaging
and the mean difference was $-0.006$  with a dispersion of 0.35 magnitudes.
The k-corrections were calculated according to the procedure
given in Lilly et al. (1995). The $(V-I)_{AB}$ color straddles 
rest-frame $B$ band of $0.2 < z < 0.9$ and the use of observed $I$-band
with an interpolation of the Coleman, Wu, \& Weedman spectral
energy distributions yields typically small corrections ($I$-band
corresponds to rest-frame B at $z\sim 0.9$) to obtain
rest-frame B magnitudes. The B-band surface brightness is derived from the
HST F814W observations and the k-corrections and rest-frame $(U-V)_{AB}$
colors are from interpolation of ground based $(V-I)_{AB}$ colors.

\subsection{Emission-line measurements}

 Measurements of the [OII] equivalent width were taken from Table 1
of Hammer et al. (1997 CFRS-XIV) for the CFRS objects, or from 
the Autofib survey (Ellis et al. 1996), or otherwise were re-measured
from the CFRS spectra themselves. The values in our Table 1 are in the 
observed frame.

\subsection{Spectroscopic failures}

 In the sample of 341 galaxies with HST imaging and spectroscopy
there are 18 objects
classified morphologically as galaxies where no redshift could be
determined from their spectra. These objects were processed along with the galaxies
with redshifts. These objects are
generally near the magnitude limit of the survey. The morphology distribution
is skewed toward early-type galaxies: 7/18 (40\%) are ellipticals
compared to 12\% Es in the entire sample. This effect is expected
because the low frequency of emission lines among early-type galaxies makes
redshift determination difficult, particularly at high redshift. 
The color distribution (see Fig. 5) of the ellipticals
without redshifts suggests that most are at $z > 0.5$. 
Lilly et al. (1995) have derived estimated redshifts for
these galaxies using the $I_{AB}$ magnitudes, the $(V-I)_{AB}$ and
$(I-K)_{AB}$ colors, and a measure of the compactness of the images
and these estimated redshifts are all at $z > 0.6$.
These estimated redshifts will be used in some of the analysis that
follows.

\section{RESULTS}

 The output from  the fitting process consists of a classification
(fractional bulge luminosity $B/T$), size (disk scale length $h$ or
bulge half-light radius $R_e$), surface brightness, 
and luminosity of each of the bulge and disk components individually.
Although colors from HST observations are available for a 
subset of the galaxies, ground-based $V-I$ colors (Lilly et al 1995, Paper I) 
are used here to ensure homogeneity.

 Table 1 shows the parameters for the sample of CFRS/LDSS galaxies that
are classified as ellipticals by the profile-fitting procedure.

\subsection{Surface brightness}

 Figure 3 shows the luminosity-size ($M_B-\log R_e$) relation for 
those galaxies classified as ellipticals by the
fitting and residual examination process. The sample is 
divided into three slices of redshift
and each panel shows the best-fit, fixed-slope, linear fit as
a solid line. The dashed line represents the local 
$M_B-\log R_e$ relation for {\em cluster} ellipticals
($M_B(AB)=-3.33 \log R_e -18.65$) derived by Schade, Barrientos, \& Lopez-Cruz (1997). 
 As described in that paper the local cluster elliptical relation
was determined from ground-based data with similar physical resolution
and signal-to-noise ratio to the HST data used here and in that
paper. The intention of using such data was to minimise 
systematic errors between our local fiducial $M_B-\log R_e$ relation
and those derived at high redshift which might occur if local data
with much higher signal-to-noise ratio and resolution were used.

Galaxies more luminous than $M_B(AB)=-20$ were used to determine
the mean shifts of the $M_B-\log R_e$ locus with ``discrepant'' points
included and excluded. ``Discrepant'' points are defined as those where the visual 
classification given in Paper I was Sab or later thus presenting
an apparent conflict with the classifications presented here based on model fits.
Duplicate observations for galaxies are treated as independent 
points but results are also calculated excluding them.
The shifts were estimated by minimising 
the residuals in magnitude (minimising the size residuals was
also tried and the difference in
the results was negligible) while
holding the slope of the linear relation fixed at the value found
by Schade, Barrientos, \& Lopez-Cruz (1997). 
The estimated shifts in luminosity are given in Table 2.

 There is significant evolution in the $M_B-\log R_e$ relation although
there is substantial scatter in the relation, particularly in the interval $0.5 < z < 0.75$.
Figure 3 uses symbols coded to indicate discrepant points (defined
above) and galaxies with measurable [OII] 3727 emission lines. 

 Elliptical galaxies with strong [OII] emission (indicating star-formation)
might be expected to be over-luminous relative to passively-evolving galaxies
if a normal initial-mass function is assumed. In the absence of that
assumption it is possible that a small number of massive, UV-bright stars
could drive the [OII] emission and blue colors without significantly
enhancing the luminosity. The observations show that they are 
roughly equally distributed on both sides of the mean  $M_B-\log R_e$ relation:
no luminosity enhancement is seen. The substantial scatter might
mask the expected effect but this issue needs further investigation with a
larger sample.

 We have also examined what effect would be produced if S0 galaxies are 
mis-classified as ellipticals. Such objects would preferentially lie to the right
of the mean $M_B-\log R_e$ relation because their disks would cause the bulge half-light radius to be
fit too large (this assumes that their bulges conform to the
passive-evolution model appropriate to the ellipticals). This is not
seen in the data.
The conclusion is that none of
these classes of objects ([OII]-strong, discrepant, possible mis-classified S0s) is
an important causal factor in the observed shifts in the $M_B-\log R_e$ relations.

 We treat spectroscopic failures by adopting their photometrically-estimated
redshift, plotting them on Figure 3 as stars, and indicating the trajectories they would
follow as a function of redshifts (from $z=0.2$ to $z=1.0$).

 The amount of evolution in the size-luminosity relation is consistent
with what is expected from passively evolving models of old
stellar populations (e.g., Bruzual \& Charlot 1993) for reasonable
values of the initial stellar mass function if the population
is sufficiently old.

\subsection{Colors}

 Figure 4 shows the color-luminosity relations for CFRS field
ellipticals (we exclude LDSS galaxies because they were observed 
in a different color system) in three redshift slices. We compute rest-frame
$(U-V)_{AB}$ as outlined in \S2.5.
The dotted line (which coincides with the solid line
in the low-redshift panel) is the adopted local
relation from Bower, Lucey, \& Ellis (1992) for Coma assuming
a redshift of 0.0232 so that $(m-M)_V=35.73$ and $(B-V)=0.96$ for ellipticals.
Noting that $(U-V)_{AB}=(U-V)+0.7$ and that $B_{AB}=B-0.17$ yields
a Coma relation $(U-V)_{AB}=-0.079M_{AB}(B)+0.51$. The colors of the ellipticals
with $0.2 < z < 0.5$ in Figure 4 agree very well with the Coma relation.

 In the following process we have assumed that the observed
colors of field ellipticals can be represented by a linear relation with
the same slope as the local {\em cluster} relation (given above) 
modified only by a uniform shift in color. In fact, a Spearman rank
correlation test does not provide significant evidence of the 
existence of a correlation between color and magnitude for the
data in Figure 4 but this is not surprising given the errors and
the shallowness of the slope of the color-magnitude relation. 

The shifts in the
color-luminosity relation are computed holding the slope fixed and
minimising the $\chi^2$ residuals in $(U-V)$. Note that the errors in color
(derived from ground-based $(V-I)_{AB}$ measurements) and luminosity 
(from HST measurements) are independent except to the extent that $(V-I)_{AB}$ was used
to compute the K-corrections. The results are given in Table 3
and the mean shifts are shown as solid lines in Figure 4.
Note that the shift $(U-V)$ for an actual galaxy requires a correction
if that galaxy evolves in luminosity as well as color. For
luminosity evolution of $\Delta M_B$ and an observed color shift
of $\Delta (U-V)$ a correction of ${{d(U-V)}\over {dM}}\times \Delta M$
is needed in the sense that a physical galaxy undergoes a smaller
blueward shift. Table 3 shifts are not corrected for this effect.

 Ellis et al. (1997) study the color-luminosity relation for
cluster E/S0 galaxies at $z\sim 0.55$. At this redshift the
$V_{555}-I_{814}$ colors approximate restframe $(U-V)$. 
The observed mean color-luminosity relation for 3 clusters
is shifted blueward by
$\simeq  -0.3$ mag in $V_{555}-I_{814}$ which is consistent with
the luminosity evolution-corrected shift of $\Delta (U-V)
=-0.26 \pm 0.08$ at $z({\rm median})=0.59$ for field ellipticals
measured here.  At higher redshifts, Stanford, Eisenhardt, \& 
Dickinson (1998) estimate 
the shift in U-infrared color to be $\simeq -0.4-0.6$ mag at $z=0.8-0.9$
for a sample of cluster early-type galaxies and we estimate a
shift of $\Delta (U-V) \simeq -0.4-0.6$ mag from Figure 1 of
Kodama, Bower, \& Bell (1998) for early-type galaxies in
the Hubble Deep Field. Our value of $\Delta (U-V)=-0.68\pm 0.11$
is consistent with those values.

 An examination of the coded symbols on Figure 4 does not suggest that
the shift in mean color is produced by
any of the sub-populations, e.g., galaxies with [OII] emission,
discrepant galaxies, or
spectroscopic failures. Although there is a hint that the ellipticals
with measurable [OII] emission are bluer than those without [OII],
the distributions in color of the two populations cannot be 
distinguished by a Kolmogorov-Smirnoff test.

A second quantity of interest, in addition to the
shift in color, is the dispersion in color ($\sigma$) at each redshift.
The observed dispersion is a combination of the intrinsic dispersion
in the color-magnitude relation and the measurement errors. We remove
the contribution from the measurement errors using a procedure from
Statistical Consulting Center for Astronomy at Pennsylvania State
University (see also Akritas \& Bershady 1996).  The observations
are denoted by $Y_i$ and their observational errors by $\sigma_i$
and the ``de-biased'' sample variance is given by $N^{-1}\sum{(Y_i-\overline{Y})}-
N^{-1}\sum{\sigma_i^2}$. An estimate of the variance of the debiassed
sample variance is $N^{-1}\sum{(\xi_i-\overline{\xi})^2}$ where $\xi_i=
Y_i^2-\overline{Y^2}-2\overline{Y}(Y_i-\overline{Y})$.

The observational errors in $(V-I)_{AB}$ were propagated through the
interpolation procedure to obtain the errors in $(U-V)_{AB}$
Given these individual errors we estimate the
intrinsic dispersion (and its 1$\sigma$ error) in the color-luminosity 
relation is $0.19\pm0.08$,  $0.29\pm0.11$ , and $0.31\pm0.16$ magnitudes in 
$(U-V)$ in the three redshift slices, respectively.
An examination of the HST imaging shows that crowding
is not a serious problem for the colors.

The dispersions in color that we find for field ellipticals 
are larger than the dispersions measured by
Bower, Lucey, \& Ellis (1992) ($\sigma < 0.06$) in the Virgo and
Coma clusters and by Ellis et al. (1997) in 3 clusters at $z\sim0.55$.
The difference is significant at roughly the $2\sigma$ level. 
 Caution requires us to note that
 a few errors in our classifications might have a large impact on
the observed dispersion. In addition, our photometric errors 
may be slightly underestimated if, for example, the photometric 
zero points vary slightly from field to field. 

On balance we conclude
that we have evidence that field ellipticals have a larger color
dispersion at a given luminosity than those in clusters, but that
the evidence is suggestive rather than conclusive.
Kodama, Bower, \& Bell (1998) examined the color-magnitude relation for
early-type galaxies with spectroscopic or photometric redshifts
in the Hubble Deep Field and concluded that half of those galaxies
are as old as rich cluster ellipticals. The intrinsic scatter found for
those galaxies with $(U-V) > 0.7$ (which is effectively the more luminous
part of the sample and is comparable to our own sample) was 
$\sigma(U-V)=0.12\pm 0.06$ compared to our somewhat 
larger value of $0.31\pm0.16$ at a similar redshift. 

\subsection{Selection of early-type galaxies: morphology versus color}

 The central tenet of this paper is that we can reliably isolate a sub-population 
of galaxies---independent of redshift---with luminosity profiles that fall into 
a particular class, namely
those whose profiles are well-fit by a deVaucouleurs $R^{1/4}$ (with no
other component required).
 For the purposes of this paper we define such galaxies as ellipticals 
provided, in addition, that they are not strongly asymmetric (asymmetry
index $R < 0.10$).
These criteria exclude
obvious S0 galaxies because they possess disk components.
Nevertheless there are borderline cases. The
E-S0 boundary is fuzzy. 

Figure 5 shows
the observed $(V-I)_{AB}$ colors versus redshift for the CFRS sample, with those
galaxies classified (from profile fitting) as ellipticals shown as filled circles.
The galaxies with fractional bulge luminosity $0.4 < B/T < 1.0$ (loosely
called S0s) are indicated by open squares.
The majority of the ellipticals are
among the reddest galaxies in the sample.  What are the other galaxies that
are very red? 
About 1/3 of the red non-ellipticals are classified as $S0$ galaxies 
(defined here as galaxies with fractional bulge luminosity $B/T \ge 0.5$),
most of the remainder are Sa-Sb galaxies (defined by $0.2\le B/T < 0.5$), 
with a few presumably reddened edge-on disks. 

 The color distribution of the elliptical sample demonstrates
that we are selecting---by morphology alone---a sample of objects with
spectral energy distributions that
are very different from the average properties of the galaxies in the full sample. 
This is encouraging.
However, the color distribution of ellipticals has large scatter and some
disk galaxies are as red as many ellipticals. This is true independent of
redshift and thus any color cut to to segregate a morphological class 
appears difficult (perhaps impossible).

 Another test of the credibility of our classifications is
a comparison with more 
conventional methods of classification. For this purpose we use the
visual classifications given for this dataset in Paper I.
In that system the early-type galaxies have classes E=0, E/S0=1, S0=2. Figure
6 shows the observed $(V-I)_{AB}$ versus redshift plot for these 
galaxies. Filled symbols are those with visual classifications
of E/S0 or earlier (numerical visual class $\le 1$ from Paper I). 
Galaxies with classifications of S0 (visual class 2) are indicated by
open squares. The visual classification system and the two-dimensional
profile-fitting system yield similar populations in terms of
color versus redshift but the dispersion in color for early-type
galaxies is large using either method of classification.

There is a fundamental difference in philosophy between selection of a
galaxy sub-population by color and
selection by morphology. The physical structure of a massive galaxy is more
likely to be a stable quantity than is the color of a galaxy. A starburst involving
a moderate fraction ($\sim 10\%$) of a galaxy's mass is capable of radically altering the
color of the galaxy (albeit temporarily). Such an event is unlikely to  
have as strong an effect on  
the apparent structure of the galaxy although admittedly this depends on the 
spatial distribution and strength of star formation. Morphological
selection keys on the apparent structure of the galaxy whereas color
selection keys on the current state of the stellar population. This current state
depends on the age distribution of its stars and the contemporaneous
rate of star-formation (as well as other quantities such as metallicity
and the initial mass function). There is also a strong dependence on dust 
content in the sense that galaxies with young or intermediate-age 
stellar populations can be shifted into the selection region by reddening.

 Kauffman, Charlot, \& White (1996) (KCW) have applied a color selection 
technique to the entire CFRS sample (the present sample is the subset of
CFRS galaxies with HST imaging). Details of their evolving stellar
population models are given in that paper.
The line plotted on figures 5 and 6 is
the selection line (intended to represent the blue edge of the
E/S0 population) applied by Kauffman, Charlot, \& White (1996) to the CFRS sample.
This line is a very poor fit to our morphologically-selected sample
but gives us the opportunity to directly compare the morphological and
color selection procedures.

Kauffman, Charlot, \& White (1996) (hereafter KCW) test the hypothesis that
giant ellipticals all formed at high redshift in a brief burst of star formation
and that thereafter they evolve passively with no further 
substantial episodes of star formation. Under this hypothesis the colors and luminosities
of those galaxies can be predicted as a function of redshift
by modelling the stellar populations.
 As a practical matter S0 galaxies are lumped together with ellipticals because their
colors are similar in the field (Buta et al. 1994) and nearly identical in
nearby clusters (Bower, Lucey, \& Ellis 1992) and clusters at higher
redshift (Ellis et al. 1997).
If the stellar models are accurate and the transformation into the observational plane is
reliable then all early-forming galaxies with no ongoing star formation can
be detected and counted.  A comparison can then be made of the space densities of
those populations at high and low redshift. If all red galaxies formed long
ago ($z >> 1$) and have since stopped forming stars then the 
space density will be conserved. If a constant space density is {\em not}
observed then either red galaxies have been created by merging since $z=1$ or those
galaxies are  bluer than the selection threshold because of recent
star formation. 

Figures 5 and 6 show that
neither the profile fitting method nor the visual classification yields a sample
that corresponds well with that selected by the color criteria
of KCW.  If a comparison is made 
with the two-dimensional profile-fitting classification system used in this paper,
color selection is both
incomplete (success rate of $< 58\%$ for E/S0s) and suffers from contamination
(47\% of the color-selected sample are not E/S0s but are other red galaxies). 
Furthermore, the
rates of completeness and contamination are a function of redshift.
For example, the percentage of ellipticals detected is 57\% at $0.2 < z < 0.5$,
36\% at $0.5 < z < 0.75$, and 21\% at $0.75 < z < 1.0$. The rates of contamination
(the fraction of galaxies redder than the threshold that are not E/S0 galaxies)
are 56\% at $0.2 < z < 0.5$, 22\% at $0.5 < z < 0.75$, and 57\% at $0.75 < z < 1.0$.
Evidently, the selection
threshold adopted by KCW does not trace
the evolution of the elliptical galaxy population as it is defined here.
 Neither does it trace the evolution of the E/S0 population as defined
by the visual classification system of Paper I. The difference is attributable
to the fact that many of our ellipticals are bluer than the KCW threshold (some
have emission lines) and many disk-dominated systems are redder than their
threshold. 
 Thus, the claim by Kauffman, Charlot, \& White (1996) that they have detected strong
evolution in the early-type population is not valid. This point will be addressed 
in section 3.6 where the $V/V_{max}$ analysis will be repeated using the 
samples selected here.

 Selection by morphology bypasses potential problems with
modelling the stellar populations, is affected very little by
moderate episodes of star formation, and is not prone to include disk galaxies 
because of the presence of dust. If the structure of these galaxies
is stable then the question is: can a set of morphological criteria be consistently applied over
a range of redshift where surface brightness dimming (a $(1+z)^4$ effect)
is a strong effect? The most likely error would be failure to detect disks
around bright elliptical components
at high redshift because of suppression by surface-brightness dimming. 
At low redshift such a galaxy would be classified as S0. At high redshift
such an object would be classified as elliptical.
In section 3.6 we will construct samples that
address these question of possible errors in the application of the
morphological selection.

\subsection{Modelling the luminosity and color evolution}

 For the moment we will ignore the presence of emission lines in
these galaxies and compare the
 evolution in color and luminosity to
Isochrone Synthesis Spectral Evolution models (Bruzual \& Charlot 1993,1996).
The GISSEL95 and GISSEL96 libraries give synthetic colors and
luminosities for simple stellar populations (and composite populations)
as a function of age with a range of metallicities. 
The available constraints
are the present-day colors of elliptical galaxies (from Buta et al. 1994), together with the
evolution in color and luminosity. If we restrict our examination to
orthodox models where the entire stellar population of elliptical  galaxies
is formed in an instantaneous burst, then the available model parameters
are the redshift of formation (assumed to be the same for all galaxies),
the metal abundance, the stellar initial mass function, and the
choice of cosmology. The choice of a particular
stellar atlas in GISSEL96 has a negligible effect on the observables in
the present case.

 Initially we adopted a Salpeter (1955) initial mass function which has been shown
(Charlot \& Bruzual 1991,Bruzual \& Charlot 1993) to reproduce the colors 
of local galaxies and we assume solar metallicity unless otherwise noted.
 We explore only cosmologies with  $H_\circ=50$ km
sec$^{-1}$ Mpc$^{-1}$ and $\lambda=0$ and use $q_\circ=0.0$ and 0.5.

 If ellipticals were a constant surface brightness population then the 
derived luminosity evolution would be independent of $q_\circ$. This is not
the case. However, the dependence on $q_\circ$ is rather weak. At $z=0.5$
using $q_\circ=0.0$ yields only $-0.09$ mag more evolution than $q_\circ=0.5$
and at $z=1$ the effect is $-0.18$ magnitudes. In the following comparisons,
these small corrections necessary for $q_\circ=0$ are applied to the 
theoretical curves rather than the data. The  major effect of varying
$q_\circ$ is to change the age of the stellar populations for a given
formation redshift.

 A single-burst population formed at $z=10$ with $q_\circ=0.5$ is shown
plotted on figure 7. The model luminosity evolution from $z=0$ to $z=1$
is somewhat larger than observed 
whereas the model colors are redder than those observed at $z \simeq 0.5$. 
Note that this plot shows both field and cluster elliptical galaxies from
previous work (reduced in a similar manner). 
The model luminosity evolution
can be reduced if $q_\circ=0.0$ (since the population is older)
but this makes the $z=0$ colors much redder and the color
evolution is then far too flat. This problem is not relieved by using either
of the other initial mass functions (Scalo 1986 or Miller and Scalo 1979)
available  in these models. Changing the metallicity fails to
produce the necessary steeper increase of color with redshift.
Since it is likely that the color systems of the models are not
identical to the observations, and because of uncertainties in
the models themselves (Charlot, Worthey, \& Bressan 1996)
differences of 2-3 tenths of a magnitude
in absolute terms will not be considered a serious disagreement. We choose
to give more weight to the comparison of differential effects with redshift.

 The observed rapid change with redshift in the mean color of the elliptical galaxy
population can be reproduced by lowering the redshift of formation.
 However, in order to obtain even a reasonable agreement between observed and model
color evolution a very recent formation epoch ($z\sim 1.5$) is required and this
enhances the luminosity evolution, creating a serious discrepancy. Another way  
for the models to reproduce the observed color evolution is
to superimpose bursts of star formation comprised of some fraction
of the galaxy mass on top of the old, passively-evolving population. Bursts
at $1 < z < 3$ with mass fractions in the range 10-25\%  are capable of
producing the color behavior but they fail to avoid the problem
of over-producing luminosity evolution; the same problem as with
a more recent, single-burst, formation epoch. A similar difficulty was noted
in Hammer et al. (1997) in fitting the 4000 angstrom break of many of the
``quiescent'' objects with an old, single-burst population.

 The fundamental problem is that a change in $(U-V)$ of $-0.68\pm 0.11$
mag should be accompanied by a much larger brightening (roughly 2 magnitudes
in the B band) in a passively evolving system.
 One particular solution to the modelling problem is to superimpose a small burst of 
star formation at $z=1$. A model that fits well is the onset at $z=1$ of
exponentially-declining star formation with an e-folding time of 1 Gyr
(a $\tau$ model) and a mass of 2.5\% of the final stellar mass of the galaxy.
Such an effect might be produced by accretion of low-mass companions
(see Silva \& Bothun 1998a) but precludes a major star-formation episode
accompanying a large disk-disk merger.
Such a model
fits very well if we adopt $q_\circ=0$ and a high redshift ($z_f=10$) of formation of
the dominant old population. We show such a model on Figure 7. (Here we have adopted
an abundance of 40\% solar for the old population and a solar abundance
for the star formation
starting at $z=1$. The difference in color at $z=0$ between the simple
models and the two-component models is due mostly to the lower abundance
chosen for the old population in these two-component models.) This model
is, of course, completely ad hoc. An attempt to provide serious constraints
on the modelling process would require accurate mean colors and luminosities
for elliptical galaxies over a wide range of redshift. 

 Nevertheless, it is evidently possible to match the observations presented
here with  simple models if the dominant stellar population is old (thus the
preference for low $q_\circ$) in order to match the slope of the luminosity
versus age relation but then some more recent ($z \simeq 1$) 
star formation is required in order to match the strong color evolution.
 Our estimates of color evolution at $z\simeq1$
are broadly consistent with both the cluster work of Stanford, Eisenhardt, \&
Dickinson (1998) and the results for field galaxies by Kodama, Bower, \& Bell (1998).
 However, precise comparisons with the present work are not possible and
we emphasise the uncertainties associated with constraining models 
with the presently very limited set of observations.

\subsection{[OII] Emission and star formation}

 In the previous section we have ignored the presence of [OII] emission lines
which indicate that these galaxies are not composed exclusively of populations
formed at $z > 1$. 
 The fraction of galaxies exhibiting strong (equivalent width $> 15$ angstroms)
[OII] is roughly 30\% independent of redshift. This compares to $< 3\%$ of a
sample of 104 nearby elliptical galaxies (Caldwell 1984). Clearly there has
been strong evolution in the emission-line properties of elliptical galaxies
since $z=1$. 

 We have estimated the star-formation rates (SFRs) for these galaxies using the
prescriptions of Kennicutt (1992) using [OII] fluxes or
equivalent widths. As emphasised in that paper there
are substantial uncertainties in such estimates. 
The SFRs derived from continuum luminosity coupled
with [OII] equivalent width compare very well with those SFRs derived
directly from the [OII] fluxes. We have adopted the former method and normalised
the SFRs in each redshift bin by the total elliptical galaxy mass in that
redshift bin (regardless of [OII] strength). The masses are estimated
from the best-fit stellar populations model in the previous section
($q_\circ=0.0$, old population plus star formation onset at $z=1$) which
gives B-band luminosity of 8.3 mag per solar mass at $z=0$. The observed
luminosities were corrected to their ``de-evolved'' luminosities at $z=0$
to yield masses.
 Galaxies with photometrically-estimated redshifts were included with
[OII] equivalent widths of zero so that they contributed to the mass
but not the star-formation rate.

 Figure 8 shows the star-formation rate per unit stellar mass 
in elliptical galaxies
as a function of redshift. It is assumed that the SFR is zero for local
ellipticals (Kennicutt 1998). If the star-formation were constant 
from $z=1$ to the present and remained at the observed high-redshift rate then
$\sim 5\%$ of the stellar mass in present-day elliptical galaxies would 
have formed since $z=1$. If, instead, the star formation rate is modelled
as exponentially declining, a smaller estimated mass fraction results.

\subsection{Merging: The density evolution of the early-type population}

 The present study shows that some level of star formation is maintained
in the elliptical population at $z < 1$. Therefore, the "orthodox"
model for the formation and evolution of elliptical galaxies is rejected.
 The remaining (very large) question is the role of late-epoch merging
in producing massive ellipticals. The role of merging can be addressed 
directly by examining the space distribution of
a complete sample of these objects. The  $V/V_{max}$ statistic (Schmidt 1968,
Lilly et al. 1995, Kauffman, Charlot, \& White 1996) 
is one way to address this question.  In the
case of no evolution $V/V_{max}$ is uniformly distributed between 0 and 1
with a mean of 0.5. The expected $1\sigma$ error in the mean of a sample of $n$ 
objects is $(12n)^{-1/2}$. We have analysed the sample of CFRS ellipticals 
(excluding LDSS ellipticals in order to produce a homogeneous sample)
 using this technique and we compare our results with those
obtained by Kauffman, Charlot, \& White (1996) using the same technique
on their color-selected subset of the whole CFRS sample.

 Table 4 presents results for the CFRS elliptical sample as well as
the early-type sample derived by Brinchmann et al. (1998) in Paper I using
visual classification techniques. We compute the mean 
value of $V/V_{max}$ and the probability ($P_{ks}$) that the
distribution of $V/V_{max}$ is drawn from a uniform parent population
(calculated using a Kolmogorov-Smirnov test).
 This directly represents the probability that a constant-space-density model is
acceptable for a specific subsample, cosmology, and assumption
about luminosity evolution as indicated in Table 4. We also
estimate the exponent $\gamma$ and its $1\sigma$ confidence interval 
for an evolution law of the functional form $F \propto (1+z)^{-\gamma}$.
The value of $\gamma$ and its error are estimated by varying $\gamma$ until 
$V/V_{max}=0.5 \pm (12n)^{-1/2}$.
 Luminosity evolution is incorporated into the analysis assuming 
that $\Delta M_B=-z$ i.e., one magnitude of evolution to $z=1$. Each galaxy
is corrected from its observed luminosity to a redshift-zero fiducial
luminosity $M_B(0)=M_B(observed)+z$. Then its apparent magnitude as a function of
redshift is computed from $M_B(z)=M_B(0)-z$ using distance modulus
and k-corrections.

 We have constructed and analysed the following subsamples. 
Sample 1 is composed of all elliptical galaxies (as judged from profile
fitting) with spectroscopic redshifts. Sample 2 adds spectroscopic failures
adopting their photometric redshifts. In sample 3 we assign all of the
spectroscopic failures a redshift of $z=0.2$. 
These samples all give results consistent with no evolution at
the 95\% significance level regardless of cosmology or luminosity
evolution.

 We address the issue of possible errors in the classifications between
E and S0 galaxies (described in the previous section) in two ways. First,
from the residuals of the fit we can make some estimate of which
ellipticals are most likely to be mis-classified. We adopt what we consider to be
an extreme level of contamination (25\%) and exclude all of these
``possible S0s'' from Sample 1. This gives us Sample 4 and the procedure has 
little affect on the $<V/V_{max}>$ results. The second way that we address
the classification error problem is to add all S0 galaxies (as judged from profile 
fitting) to Sample 1 to yield Sample 5. All of these results are consistent
with no-evolution models of the population. Note that if we were to add
spectroscopic failures at their photometric redshifts we would make
$<V/V_{max}>$ larger whereas $<V/V_{max}>$ needs to be smaller than $0.5$ 
in order to produce the Kauffman, Charlot, \& White (1996) result. 

 An alternative to using the classifications given by the model-fitting approach
is to adopt the visual classifications from Paper I.
 Samples 6 and 7 are constructed using these classifications with and without
photometric redshifts, respectively. Sample 6 results in rejection of the
no-evolution hypothesis at the 95\% significance level for $q_0=0.01$. Sample
7 is preferred over sample 6 because it includes spectroscopic failures at
their photometrically-estimated redshifts. Sample 7 yields 
a result consistent with no-evolution models.

The results in Table 4 provide no evidence 
of evolution in the space
density of early-type galaxies regardless of whether that population
is defined as elliptical galaxies (by the profile-fitting method) or
as E/S0 galaxies using the visual classifications. The most reliable 
samples in Table 4 are those where photometric redshifts have been 
adopted where spectroscopic redshifts were not available. These samples
are the most reliable because they correct for the expected incompleteness
in spectroscopic redshifts for early-type galaxies at high redshift.
Considering those ``best''
samples indicates that no-evolution models are consistent with these
datasets independent of cosmology and independent of luminosity evolution.

 This conclusion about the evolution of ellipticals from the $V/V_{max}$ statistic
contradicts that reached by Kauffman, Charlot, \& White (1996). The color selection
method employed in that paper fails to produce a complete, uncontaminated
sample of early-type galaxies. Totani \& Yoshii (1998) employ their own models
to select early-type galaxies on the basis of color. They apply a sample cut-off
at $z=0.8$ based on the assumption that the CFRS is not complete at higher
redshifts. With this cut-off they find $<V/V_{max}>=0.478\pm0.035$ which is
consistent with no evolution. 

It is important to
appreciate that the size of the error bars allow a factor of 2-3 change
in the space density at the $1\sigma$ level. Therefore, although the present work
provides no evidence for significant evolution in the space density of early-type galaxies
{\em it does not rule out such evolution}. A larger sample would be required to
decide the question using a $<V/V_{max}>$ test.

 Finally, we can compare the observed space density of elliptical galaxies in
the CFRS sample to the predictions based on the local luminosity function. For this
purpose we use the E/S0 LF of Marzke et al. (1998) (a conversion is required
from their $H_\circ=100$ km sec$^{-1}$ Mpc$^{-1}$ to our value of
$H_\circ=50$ km sec$^{-1}$ Mpc$^{-1}$). The LF is integrated over all galaxies
brighter than $M_{B}(AB)=-20.2$ to yield a space density of $3.47\times 10^{-3}$
Mpc$^{-1}$. The effective area of the HST subsample of the CFRS is taken to
be 0.0096 deg$^2$. The expected numbers are computed for the three redshift shells
$0.2 < z < 0.5$, $0.5 < z < 0.75$, $0.75 < z < 1.0$.
In the present case a negligible error is introduced by assuming
that each galaxy can be detected throughout is redshift shell and
simply multiplying the integral of the luminosity function above by the
total volume within the shell, given the area of coverage.
We assume that the luminosity evolution of
the elliptical population is $M(z)=M(z=0)-z$ which approximates the results
of the present study. Thus we count galaxies in $0.2 < z < 0.5$ brighter than 
$M_{B}(AB)=-20.2$, those in $0.5 < z < 0.75$ brighter than $-20.7$, and
galaxies in $0.75 < z < 1.0$ brighter than $M_{B}(AB)=-21.2$. 

The comparison between observed and predicted numbers are given in Table 5. Because
we are comparing classifications derived from visual inspection with those
derived from profile-fitting (and at very different redshifts), there is the 
potential for systematic differences in the populations being counted. On the 
other hand, we have shown that the visual classifications done on this sample
agree broadly with the profile-fitting classifications. A puzzling feature
of the result shown in Table 5 is that we count roughly the same number of
elliptical galaxies as are predicted for $q_\circ=0.01$ by the LF for
elliptical and S0 galaxies combined. Data from Buta et al. (1994) suggests
that S0 galaxies will comprise roughly two-thirds of the E/S0 sample whereas
the LF's from Marzke et al. 1994 suggest that only one-quarter of an
E/S0 sample will consist of ellipticals.

 Table 5 provides no evidence for a deficit of elliptical galaxies at high
redshift. On the contrary, the problem is that there are more than 
expected if ellipticals
constitute only a fraction (1/3 to 1/4) of the galaxies in the E/S0 luminosity
functions used in the calculations. The problem is substantially worse in the
case where $q_\circ=0.5$. However, if isolated ellipticals gradually accrete
halo gas to build a disk (as suggested in some formation scenarios) then
these high-redshift ellipticals might evolve into present-day S0 galaxies. 
 An alternative is that we have failed to detect the faint disks around
some early-type galaxies at high redshift that would a have resulted in
a classification of S0 rather than E. Either of these interpretations 
would result in perfect agreement with the numbers in Table 5 if
$q_\circ\sim0$.

\section{CONCLUSIONS}

 The analysis of the CFRS/LDSS sample of field elliptical galaxies reveals
evolution in the $M_B-\log R_e$ relation so that a galaxy of a given
size is more luminous by $-0.97\pm 0.14$ magnitudes at $z=0.9$ than its
local counterpart as estimated from the cluster elliptical locus.
 This apparent evolution in  luminosity is accompanied by
a bluing of these galaxies with look-back time. At $z=0.9$ the mean
shift in color is $-0.68 \pm 0.11$ mag in rest-frame (U-V). 
 The luminosity evolution is consistent with simple models of passive evolution
of an old, single-burst stellar population but a small quantity
(by mass) of more recent star formation is required to make the models match the
observed color evolution. Approximately 1/3 of the elliptical
galaxies at $0.2 < z < 1.0$ display [OII] emission lines, indicating
star formation at rates that we estimate at $\sim 0.5 $M$_{\odot}$ yr$^{-1}$ 
per $10^{11}$ M$_{\odot}$ of stellar mass in elliptical galaxies so that
perhaps a few percent of the stellar populations of these galaxies have
formed since $z=1$.

 The estimates of luminosity and color evolution found here are
consistent with that found in clusters (Schade, Barrientos, \& L\'{o}pez-Cruz 1997,
Stanford, Eisenhardt, \& Dickinson 1998). 
On the other hand, we have found evidence that the dispersion
in the color-luminosity relation may be larger among field
ellipticals at $z > 0.5$ than the dispersion observed for early-type galaxies in Coma
by Bower, Lucey, \& Ellis (1992) or those in clusters at $z=0.55$ by
Ellis et al. (1997). 

 The presence of [OII] emission indicates that star formation is occurring
in $\sim 30\%$ of the elliptical galaxies in this sample. Thus the "orthodox"
view of elliptical galaxy formation (a single burst followed by no further
star formation) is false. A larger dispersion in color among field ellipticals than
among those in clusters (which is suggested by these data)
would also argue in favor of a more diverse star formation history in field
ellipticals than in cluster early-type galaxies in the sense that more
recent activity occurred in field galaxies possibly because continued gas infall
that could fuel ongoing star formation is suppressed in the cluster environment.

 We have repeated the $<V/V{max}>$ test for early-type CFRS galaxies
done by Kauffman, Charlot, \& White (1996) using a subset of the
CFRS sample with HST imaging. Our sample was selected on the basis
of profile fitting rather than color (the technique they used).
We find no evidence  for evolution in the space density of large ellipticals
over the range $0.2 < z < 1.0$ within the CFRS survey.
We cannot rule out fairly large changes (factors $\sim$ 2-3) in
density because of the small numbers in our survey. Nevertheless,
our result contradicts the claim for detection of strong evolution
by Kauffman, Charlot, \& White (1996) because the
color selection applied in that work did not detect all of the E/S0 galaxies
that were present and also suffered from contamination by red disk-dominated
galaxies.  

A more powerful test for a change in space density (although more prone to systematic 
error because of differences in the classification method)
is a comparison of our observed number of field elliptical galaxies at
high redshift with the predictions based on the E/S0 luminosity function
of Marzke et al. (1998). This test indicates no deficit of early=type galaxies at
$z \le 1$. In fact, there is a surplus of ellipticals at high redshift unless
the ratio of E to S0 galaxies increases at high redshift. Such an effect might
be spurious. It is plausible that faint disks might be missed at high
redshift resulting in mis-classification of S0 galaxies as ellipticals.
Alternatively, it might be a real evolutionary effect where ellipticals
are transformed into S0s via the development of disk components. Formally, we
can say that, at the 95\% confidence level, the number of ellipticals that
we detect at $0.75 < z < 1.0$ is between $0.71$ and $2.03$ times the number
of E/S0 galaxies predicted by the Marzke et al. (1998) luminosity function assuming
$q_\circ=0.01$ and including spectroscopic failures. 

 In summary, the results presented here are consistent with an early formation
epoch for these elliptical galaxies but with some degree of ongoing star formation
at $z < 1$. We reject the orthodox model of elliptical galaxy evolution.
We find no evidence that the space density of large ellipticals has
changed since $z=1$. The result of Im et al. (1996) agrees very well with
our result and carries substantially more statistical weight. These results
directly challenge the view that significant numbers of elliptical galaxies have
been formed by mergers since $z=1$.

 A final note on the question of merging. Le F\`evre (1999) find
evidence---from the same sample of galaxies studied here---that the
rate of bright galaxy mergers increases steeply with redshift. That
work suggests that a typical ($L^*$ at $z\sim 1$ ) galaxy will undergo 1-2 major merging
events since $z=1$. On the other hand, the present study finds no evidence
for a significant change in the space density of massive
elliptical galaxies since $z=1$. Furthermore, Lilly et al. (1998) find
no significant change in the space density of large disk galaxies
over this redshift range. These observations can be reconciled 
if the enhanced level of interaction and merging that is seen is, in fact,
taking place
among galaxies that are somewhat less massive than present-day $L^*$
galaxies and if such interactions do not frequently produce large elliptical
galaxies. 

\begin{acknowledgments}

We acknowledge the support of NATO in the form of a travel grant.
The research of SJL was supported by NSERC. 

\end{acknowledgments}

\newpage
\centerline{Figure Captions}

\begin{figure}
\caption{ Differences in best-fit parameters from multiple observations
of galaxies. The bottom panel shows the fractional bulge luminosity, $B/T$,
in the center is the total magnitude, and on the top is the
index of asymmetry.  In the bottom and center panels, filled circles represent 
symmetric, or morphologically normal galaxies whereas open circles show
asymmetric galaxies. The $B/T$ value is not considered a valid classification
index for asymmetric galaxies. The errors in a single measurement
(for symmetric galaxies) as inferred from these duplicate measurements are
shown on each panel. The right-most panel indicates the cut-off in
asymmetry index adopted here (solid line) and in Schade et al. (1995).}

\end{figure}

\begin{figure}
\caption{Differences in structural parameters from mutltiple observations. 
Filled symbols represent symmetric (normal) galaxies and open symbols indicate asymmetric galaxies.
 The values of $\sigma$ (computed excluding asymmetric galaxies) are shown
as inferred for a single measurement from
the analysis of these duplicate fits.}

\end{figure}

\begin{figure}
\caption{ The relation between $M_{B}(AB)$ and log of the half-light
radius ($\log R_e$)
for elliptical galaxies in three slices of redshift for field
galaxies in the CFRS/LDSS survey. The dashed line in
each panels corresponds to the best-fit to the local sequence
and the solid lines indicate the best-fit {\it fixed-slope}
($\Delta M_B/\Delta \log R_e=-3.33$) relation for each slice. In 
all cases the solid lines are the best-fits with all points
included. Open symbols are `` discrepant'' galaxies defined as those
objects where visual classification
gives a type earlier than S0 in apparent conflict with the class E derived
by profile fitting. Spectroscopic failures (galaxies without spectroscopic
redshifts) are plotted as stars at their photometrically-estimated 
redshifts and their trajectories as a function of redshift are indicated by
light dashed lines. Large open circles indicate galaxies which have
measurable [OII] emission lines. There is no clear indication that
any of these sub-classes (``discrepant'',[OII]-emission, spectroscopic failures) 
is responsible for the shift in the mean $M_{B}(AB)-\log R_e$ locus.}

\end{figure}

\begin{figure}
\caption{ The color-luminosity relations for elliptical galaxies in
three redshift slices. Dashed lines denote the Coma relation derived
from Bower, Lucey, \& Ellis (1992) and the solid lines are the best-fit
relations (derived holding the slope of the relation
fixed at the Coma value) for the CFRS galaxies. Symbols and loci are as defined
for figure 3. As in figure 3 there is no evidence that any of the sub-classes
is responsible for the evolving mean color locus. Such an effect might be 
difficult to detect in our small sample.}

\end{figure}

\begin{figure}
\caption{ Observed colors versus redshift for the CFRS galaxies with
HST imaging.
Filled symbols are those galaxies classified as ellipticals 
and open squares indicate galaxies classified as S0's 
(defined here as those with fractional bulge luminosities $0.4 < B/T < 1.0$)
by
the fitting process. Asymmetric objects have been excluded.
Spectroscopic failures are plotted as filled symbols at $z=0$ if they
have been classified as ellipticals and as open symbols at $z=1.4$
if they are later types. The color-selection criterion defined by Kauffman,
Charlot, \& White 1996 (plotted as a solid line) selects only a small subset
of those galaxies that are classified as ellipticals or S0's by
the profile fitting process.}
\end{figure}

\begin{figure}
\caption{Same as in figure 5 except that the visual classifications
of Brinchmann et al. (1998) are used instead of the classifications from
the fitting procedure. Again the Kauffman,Charlot, \& White 1996 selection
line selects only a subset of the galaxies classified visually in Paper I
as early-type galaxies. }
\end{figure}

\begin{figure}
\caption {The derived shifts in luminosity (upper panels); filled circles 
are field ellipticals from the present work, filled pentagons are field
ellipticals from Schade et al. (1996); open circles are cluster ellipticals
from Schade, Barrientos, \& L\'{o}pez-Cruz 1997, and Schade et al. (1996).
Lower panels are shifts in color from the present work. 
The colors
have been corrected to represent the evolution in color of a physical galaxy
(which evolves in color {\em and} luminosity) rather than the evolution of
the mean color relation. The solid lines
represent a model consisting of a single-burst population 
formed at $z=10$ and the dotted lines are models
where there is an added episode of exponentially declining star formation 
comprised of
2.5\% of the final stellar mass of the galaxy beginning at $z=1$. The single-burst
models have solar abundances and the two-component models use an
abundance of 40\% solar for the old population and a solar abundance for
the younger population. These abundance differences are the main cause
of the difference in one-component and two-component model colors at $z=0$.}
\end{figure}

\begin{figure}
\caption {The rate of star formation per unit stellar mass in elliptical
galaxies as a function of redshift. At $z=0$ the star-formation rate is
taken to be zero (Kennicutt 1998). The star-formation rate is estimated
from the equivalent width of [OII] emission lines and the mass is estimated 
from model mass-to-light ratios at $z=0$ and galaxy luminosities that have
been corrected (to $z=0$ values) for luminosity evolution. Galaxies without
[OII] emission have been included in the mass estimates.}
\end{figure}


\newpage

\begin{references}{}

\reference{} Abraham, R., Ellis, R., Fabian, A., Tanvir, N., \& Glazebrook, K. 
1998 astro-ph 9807140

\reference{} Akritas, M. \& Bershady, M. 1996 \apj 740, 706

\reference{} Aragon-Salamanca,A., Ellis,R., Couch,W., \& Carter,D. 1993, \mnras, 262 764

 \reference{} Bender, R. 1992 \aa, 258, 250

\reference{} Barger, A., Cowie, L., Trentham, N., Fulton, E., Hu, E., Songaila, A.,
\& Hall, D. 1998 astroph 9809299

\reference{} Barnes, J. \& Hernquist, L. 1996 \apj, 471, 115 

\reference{} Barrientos,F., Schade, D., \& L\'{o}pez-Cruz, O. 1996 \apjl, 460, 89 

\reference{} Baugh, C., Cole, S., \& Frenk, C. 1996 \mnras 283, 1361

\reference{} Bender,R. Ziegler, B., \& Bruzual, G. 1996, \apjl 463, 51 

\reference{} Bower, R., Kodama, T., \& Terlevich, A. 1998, \mnras, 299, 1193

\reference{} Bower, R., Lucey, J., \& Ellis, R. 1992 \mnras 254, 601

\reference{} Brinchmann, J., Abraham, R., Schade, D., Tresse, L. et al. 1998 \apj 499, 112

\reference{} Bruzual, G. \& Charlot 1993, \apj, 405, 538

\reference{} Buzzoni, A., 1995, \apjs, 98, 69

\reference{} Burstein, D. \& Heiles, C. 1982, \aj, 887, 1165

\reference{} Buta, R., Mitra, S., deVaucouleurs, G., \& Corwin, H. 1994, \aj, 107, 118

\reference{} Capelato, H., de Carvalho, R., \& Carlberg, R. 1995, \apj, 451, 525
 \apj, 457, 625

\reference{} Carlberg, R., Pritchet, C., \& Infante, L. 1994, \apj, 435, 540

\reference{} Charlot, S. \& Bruzual, G. 1991, \apj, 367, 126

\reference{} Charlot, S., Worthey, G. \& Bressan, A. 1996, \apj, 457, 625

\reference{} Coleman, G., Wu, C. \& Weedman, D. 1980, \apjs, 43, 393

\reference{} Crampton, D., LeF\`evre, O., Lilly, S., \& Hammer, F. 1995, ApJ, 455, 96

\reference{} de Carvalho, R., \& Djorgovski, S., 1992, \apjl, 389,49

\reference{} Dickinson, M. 1995 in {\em Fresh Views of Elliptical
Galaxies}, A.S.P. Conference Series, ed. Buzzoni,A., Renzini, A., \&
Serrano, A. p. 283

\reference{} van Dokkum, P., Franx, M., Kelson, D., \& Illingworth, G. 1996, \apj, 504, L17 

\reference{} van Dokkum, P. \& Franx, M. 1996, \mnras, 281, 985 

\reference{} Djorgovski, S., \& Davis, M. 1987, \apj, 313 59

\reference{} Dressler, A., Lynden-Bell, D., Burstein, D., Davies, R., Faber, S.,
Terleveich, R., Wegner, G. 1987, \apj, 313, 42

\reference{} Dressler, A., \& Gunn, J. 1990, in {\it Evolution of the universe of galaxies
}, San Francisco,Astronomical Society of the Pacific

\reference{} Driver, S., Windhorst, R., Phillips, S., \& Bristow, P. 1996, \apj, 525, 533

\reference{} Driver, S., Fernandez-Soto, A., Couch, W., Odewahn, S., Windhorst, R.,
Phillips, S., Lanzetta, K., \& Yahil, A. 1998 \apj, 496, 93

\reference{} Eggen, O., Lynden-Bell, D., \& Sandage, A., 1962 \apj, 136, 748

\reference{}  Ellis, R., Smail, I., Dressler, A., Couch, W., Oemler, A., Butcher, H.,
\& Sharples, R. 1997 \apj, 483, 582  

\reference{} Ellis, R., Colless, M., Broadhurst, T., Heyl, J., \& Glazebrook, K.
1996, \mnras, 280, 235

\reference{} Elmegreen, B., Elmegreen, D., \& Montenegro, L. 1992, \apjs, 79, 37

\reference{} Feigelson, E., \& Babu, G. 1992 \apj, 397, 55

\reference{} Forbes, D., Ponman, T. \& Brown, R. 1998 \apjl, 508, 43

\reference{} Franceschini,A., Silva,L., Fasano, G., Granato, G., Bressan, A.,
Arnouts, S., \& Danese, L. 1998, \apjl, 506, 600

\reference{} Freeman, K. 1970 \apj 160, 811

\reference{} Guzman, R., \& Lucey, J. 1993 \mnras 263, 47

\reference{} Hammer, F. et al. 1997, \apj, 481, 49

\reference{} Hau, G. \& Thompson, R. 1994 \mnras 270, 23

\reference{} Im, M., Griffiths, R., Ratnatunga, K., Sarajedini, V. 1996, \apj, 461, 79

\reference{} de Jong, R. 1996 \aaps, 118, 557

\reference{} Jorgensen, I., Franx, M., \& Kjaergaard, P. 1996, \mnras, 280, 167

\reference{} Kauffmann, G. \& Charlot, S., 1998, \mnras, 294, 705

\reference{} Kauffmann, G. 1996 \mnras, 281, 487

\reference{} Kauffmann, G., Charlot, S., \& White, S. 1996 \mnras, 283, 117

\reference{} Kauffman, G., White, S., \& Guiderdoni, B. 1993 \mnras, 264, 201

\reference{} Kelson, D., van Dokkum, P., Franx, M., Illingworth, G.,
Fabricant, D. 1997 \apjl, 478, 13

\reference{} Kennicutt, R. 1992 \apj 388, 310

\reference{} Kennicutt, R. 1998  astroph/9807187

\reference{} Kent, S. 1985 \apjs 59, 115

\reference{} Kodaira,K. Watanabe,M. \& Okamura,S.
     1986, \apjs, 62, 703

\reference{} Kodama, T., Bower, R., \& Bell, E. 1998 astroph 9810138

\reference{} Kormendy, J. 1977, \apj, 218, 333

\reference{} van der Kruit, P. 1987 \aap 173,59

\reference{} Larson, R. 1975, \mnras 173, 671

\reference{} Le F\`evre, O., Abraham, R., Lilly, S., Ellis, R., Schade, D.,
Tresse, L., Colless, M., Crampton, D., Glazebrook, K., Hammer, F., \&
Broadhurst, T. 1999, \mnras submitted.

\reference{} Lilly, S.J., Tresse, L., Hammer, F., Crampton, D. \& Le F\`evre, O. 1995,
\apj, 455, 108

\reference{} Lilly, S.J., Le F\`evre, O., Crampton, D., Tresse, L., 1995 \apj, 455, 50

\reference{} Lilly, S., Schade, D., Ellis, R., Le F\`evre, O. et al. 1998 \apj 500, 75

\reference{} L\'{o}pez-Cruz, O. 1996, PhD Thesis, University of Toronto

\reference{} Marzke, R., Da Costa, L., Pellegrini, P., Willmer, N., \& Geller, M.
1998, \apj, 503, 617

\reference{} Marzke, R., Geller, M., Huchra, J., \& Corwin, H. 1994 \aj, 108, 437

\reference{} Menanteau, F., Ellis, R., Abraham, R., barger, A., \& Cowie, L. 1998 astroph/9811465

\reference{} Mihalas, D., \& Binney, J. 1981 in  {\it Galactic Astronomy},
San Francisco, W.H. Freeman and Company, p. 189

\reference{} Miller, G., \& Scalo, J. 1979, \apjs 41, 513

\reference{} Mobasher, B. \& James, P. 1996, \mnras, 280, 902

\reference{} Negroponte, J. \& White, S. 1983, \mnras, 205, 1009 

\reference{} Oke, J., Gunn, J., \& Hoessel, J. 1996, \aj, 111, 29

\reference{} Patton, D., Pritchet, C., Yee, H., Ellingson, E., \& Carlberg, R.
1997, \apj, 475, 29

\reference{} Pahre, M., Djorgovski, S., \& de Carvalho, R. 1998, \aj, 116, 1606

\reference{} Pahre, M., Djorgovski, S., \& de Carvalho, R. 1996, \apjl, 456, 79

\reference{} Press,W.H., Teukolsky,S.A.,  Vetterling,W.T., \& Flannery,B.P.
 1992,  Numerical Recipes (Cambridge University Press: Cambridge)

\reference{} Rakos,K., \& Schombert,J. 1995, \apj, 439, 47

\reference{} Salpeter, E. 1955, \apj, 121 161

\reference{} Scalo, J. 1986 Fund. Cosmic Phys. 11, 1

\reference{} Sandage, A. \& Visvanathan, N. 1978 \apj, 225, 742

\reference{} Schade, D., Barrientos, L. F., \& L\'{o}pez-Cruz, O. 1997
\apjl 477, 17

\reference{} Schade, D., Carlberg, R., Yee, H., L\'{o}pez-Cruz, O., \&
Ellingson, E. 1996a, \apjl, 464, 63  

\reference{} Schade, D., Lilly, S., Crampton, D., Le F\`evre, O., Hammer, F.,  
\& Tresse, L. 1995, \apjl, 451, 1

\reference{} Schombert, J. \& Bothun, G. 1987 \aj, 92, 60

\reference{} Schmidt, M. 1968 \apj, 151, 393

\reference{} Schweizer, F., \& Seitzer, P. 1992 \aj 104, 1039

\reference{} Silva,D. \& Bothun, G. 1998a \apj,  116, 85

\reference{} Silva,D. \& Bothun, G. 1998b \aj,  116, 2793

\reference{} Simien, F. \& de Vaucouleurs, G. 1986, \apj, 302, 564

\reference{} Stanford, S.A., Peter R. Eisenhardt, P., \& Dickinson, M. 1998 \apj
492, 461 

\reference{} Stetson, P.B., 1987, \pasp, 99, 191

\reference{} Tinsley, B. 1972, \apj, 178,319

\reference{} de Vaucouleurs, G.
     1948, Ann. d'Astrophys. 11, 247

\reference{} White, S. \& Rees, M. 1978, \mnras, 183, 341

\reference{} Whitmore, B. 1995 {\em Photometry with the WFPC2} available on
Space Telescope Science Institute WWW pages.

\reference{} Yee, H.K.C. 1991, \pasp, 103, 396

\reference{} Zepf, S., 1997 Nature, 390, 377

\reference{} Zepf, S., \& Koo, D., 1989 \apj, 337, 34

\end{references}
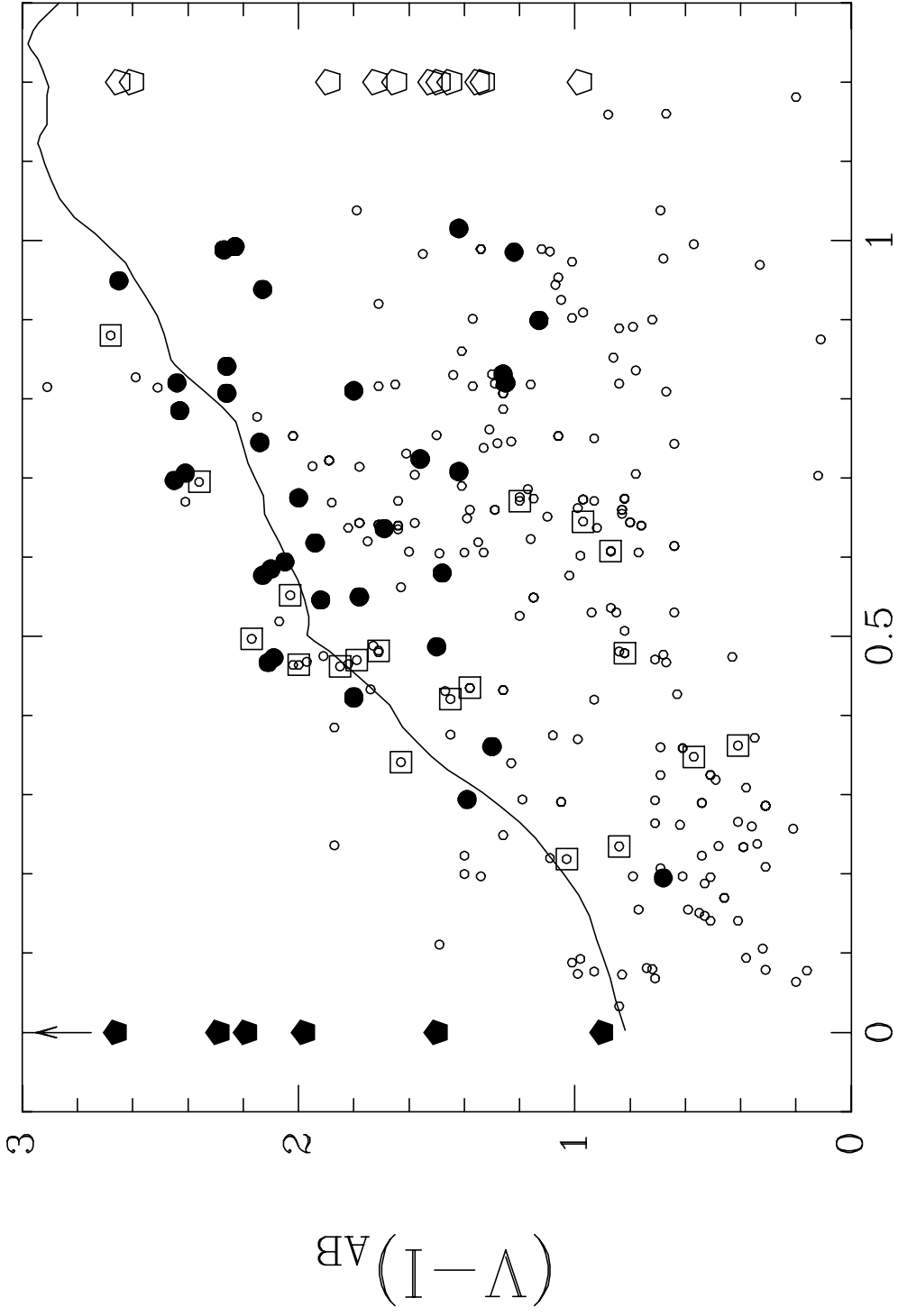
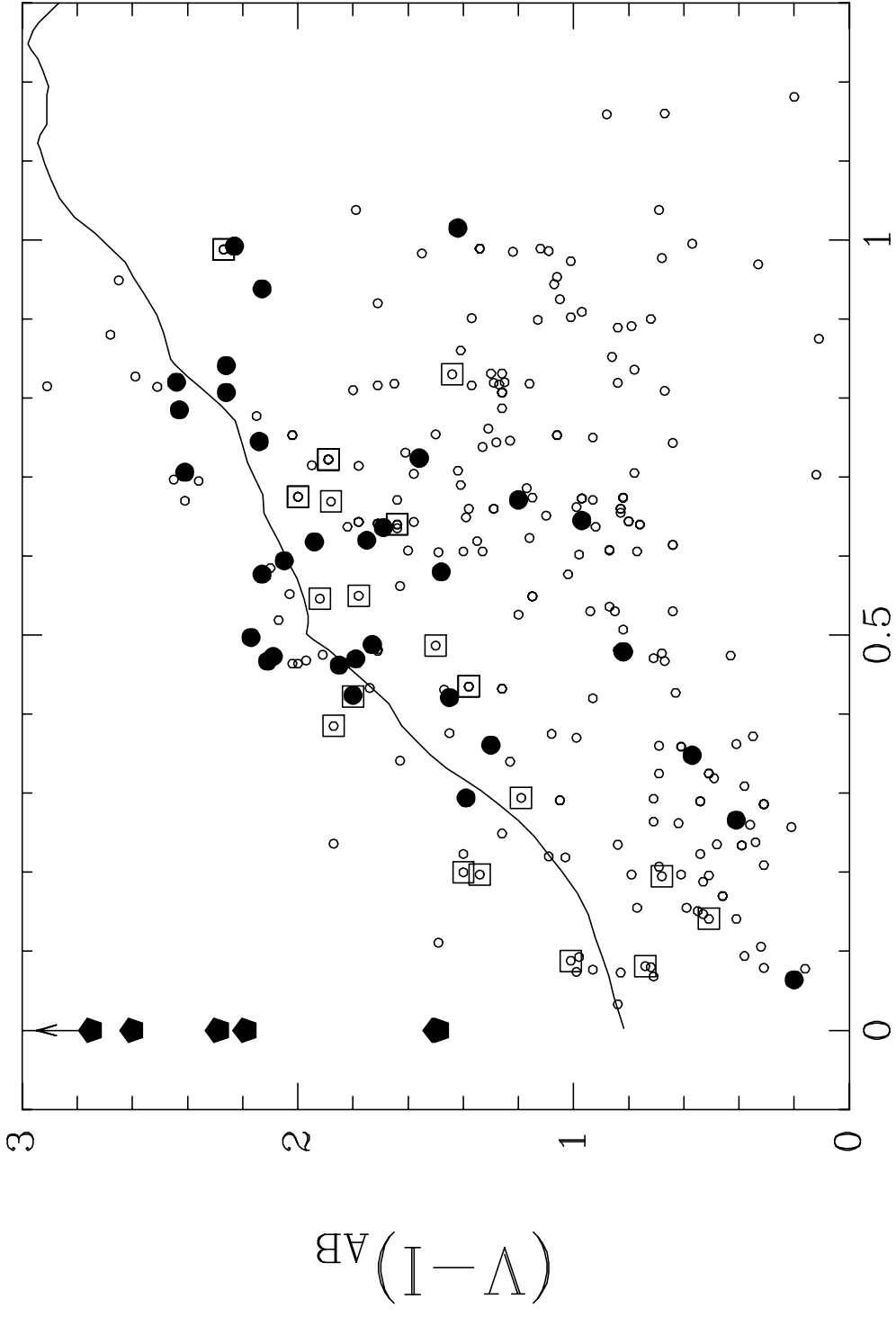
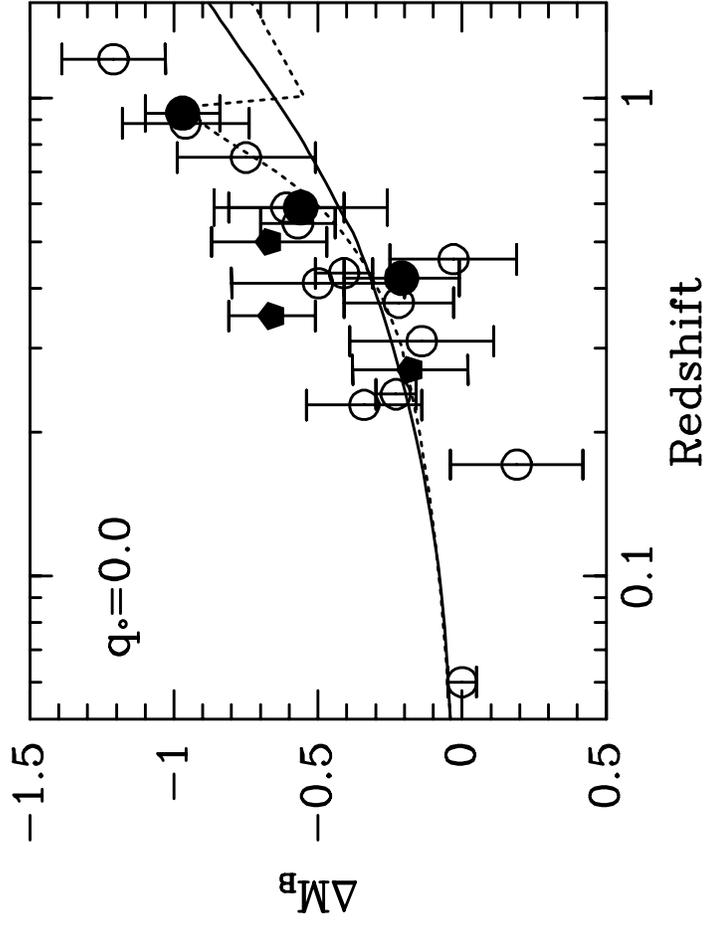
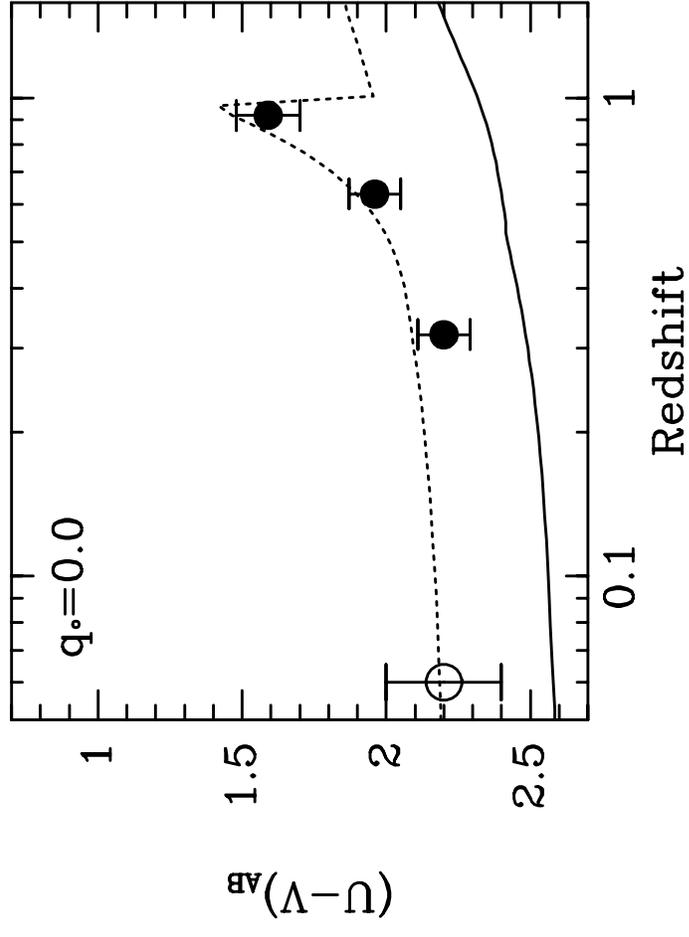
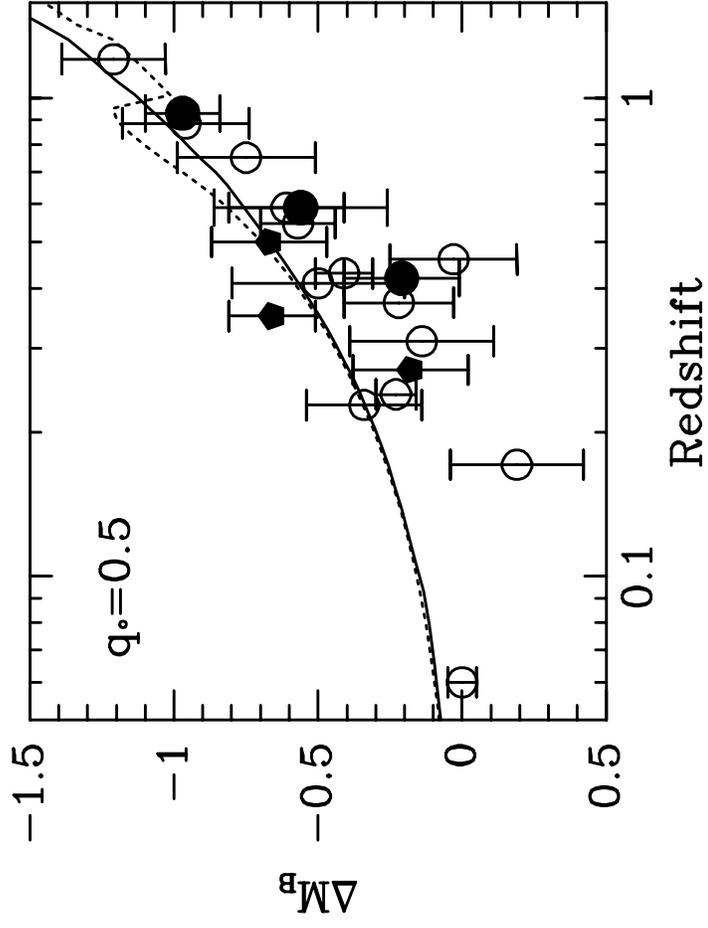
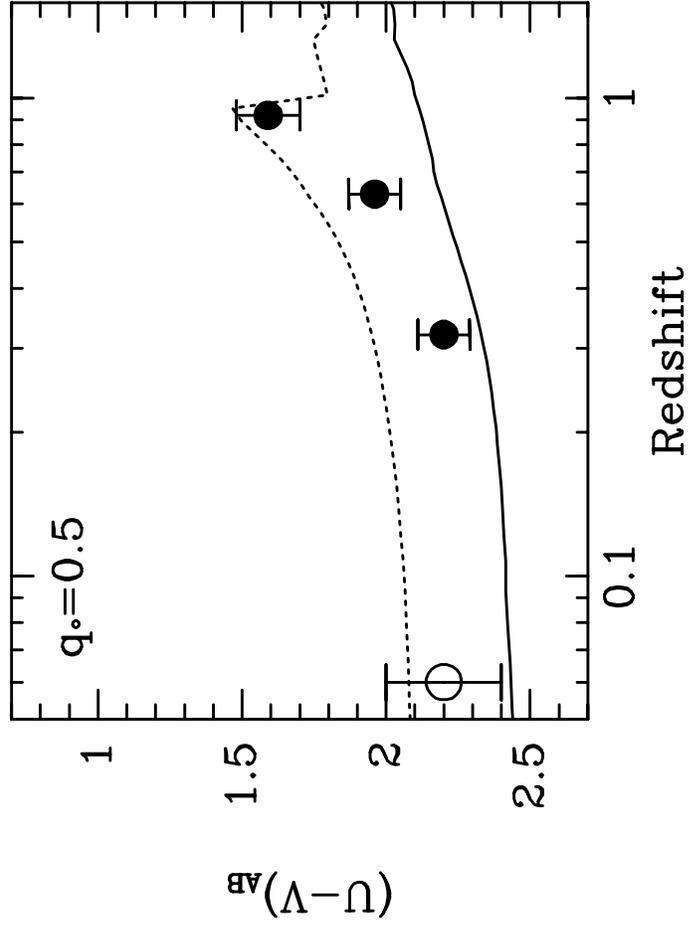
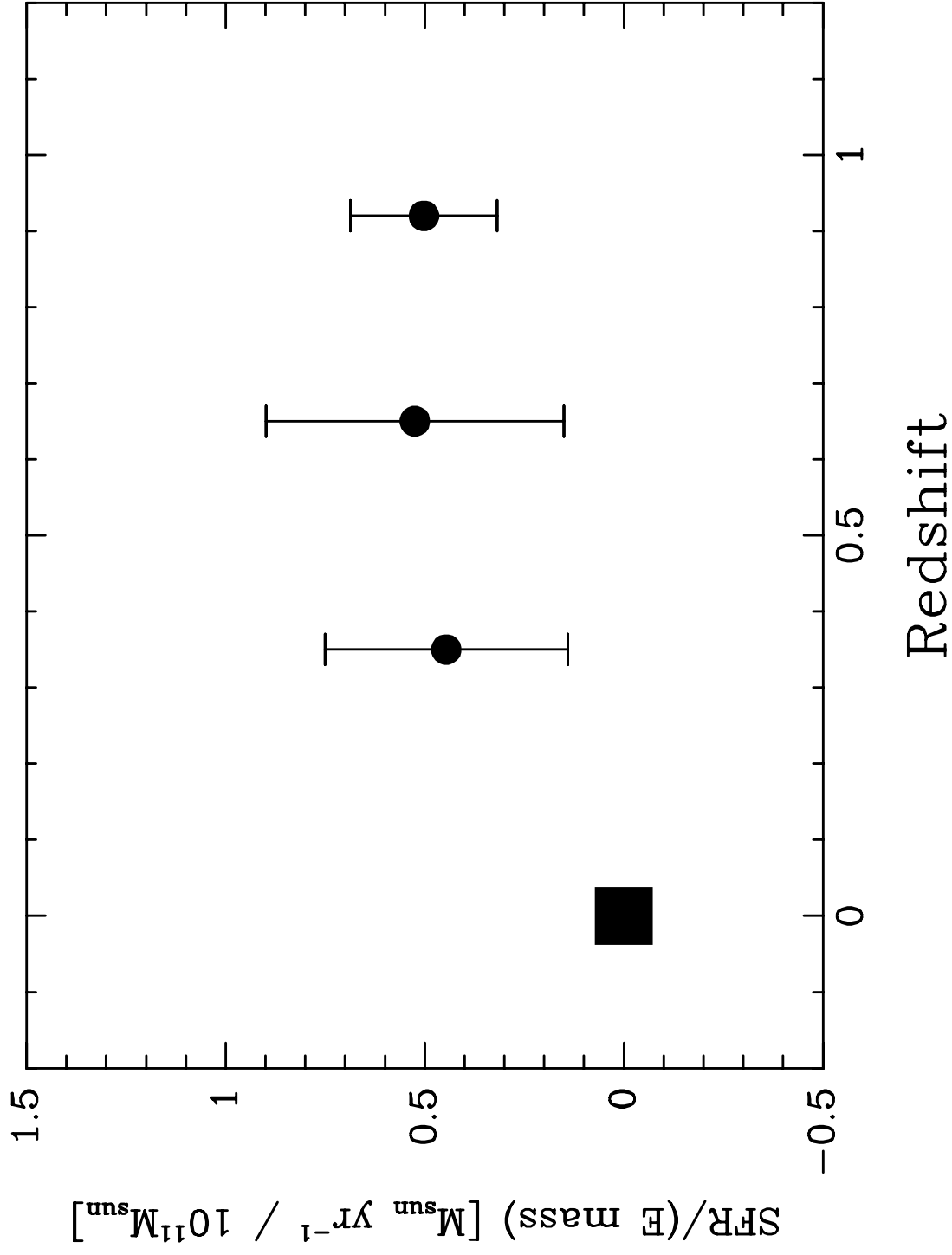
\end{document}


\begin{deluxetable}{llrlll}
\tablenum{4}
\tablewidth{0pt}
\tablecaption{Results of the $V/V_{max}$ test}
\tablehead{
\colhead{Sample} & \colhead{$q_\circ$} &
\colhead{LE} & \colhead{$<V/V_{max}>$} & \colhead{$P_{ks}$} & \colhead{$\gamma$} }
\startdata
Sample 1              & $0.5$    & NO   & $0.57\pm 0.05$ & 0.30  & -2.4 [-4.0,-0.8] \nl 
Sample 1              & $0.01$   & NO   & $0.54\pm 0.05$ & 0.50  & -1.5 [-2.9,0.1] \nl 
Sample 1              & $0.5$    & YES  & $0.55\pm 0.05$ & 0.42 & -1.5 [-3.1,-0.1] \nl  
Sample 1              & $0.01$   & YES  & $0.52\pm 0.05$ & 0.86 & -0.7 [-2.3,0.8] \nl  
 & &  \nl    
Sample 2           & $0.5$    & NO   & $0.61\pm 0.05$ & 0.07 & -3.9 [-5.7,-2.3] \nl 
Sample 2           & $0.01$   & NO   & $0.58\pm 0.05$ & 0.20 & -3.1  [-5.0,-1.5] \nl 
Sample 2           & $0.5$    & YES  & $0.59\pm 0.05$ & 0.12 & -2.9 [-4.5,-1.4] \nl 
Sample 2           & $0.01$   & YES  & $0.56\pm 0.05$ & 0.57 & -2.1 [-3.8,-0.6] \nl
 & &  \nl    
Sample 3           & $0.5$    & NO   & $0.47\pm 0.05$ & 0.18  & 0.7 [-0.9,2.3] \nl 
Sample 3           & $0.01$    & NO   & $0.45\pm 0.05$ & 0.18 & 1.7 [0.0,3.3] \nl 
Sample 3           & $0.5$    & YES   & $0.46\pm 0.05$ & 0.18 & 1.3 [-0.1,2.8] \nl 
Sample 3           & $0.01$    & YES   & $0.43\pm 0.05$ & 0.18 & 2.2 [0.7,3.8] \nl 
 & &  \nl    
Sample 4           & $0.5$   & NO   & $0.59\pm 0.06$ & 0.09 & -3.0  [-5.0,-1.2] \nl 
Sample 4           & $0.01$   & NO   & $0.56\pm 0.06$ & 0.13 & -2.2  [-4.3,-0.4] \nl 
Sample 4           & $0.5$   & YES   & $0.58\pm 0.06$ & 0.19 & -2.1  [-3.9,-0.4] \nl 
Sample 4           & $0.01$   & YES   & $0.54\pm 0.06$ & 0.36 & -1.3  [-3.2,0.5] \nl 
 & &  \nl    
Sample 5           & $0.5$   & NO   & $0.49\pm 0.04$ & 0.75 & 0.1  [-1.2,1.2] \nl 
Sample 5           & $0.01$   & NO   & $0.47\pm 0.04$ & 0.36 & 0.9  [-0.4,2.1] \nl 
Sample 5           & $0.5$   & YES   & $0.47\pm 0.04$ & 0.50 & 1.7   [-0.5,2.9] \nl 
Sample 5           & $0.01$   & YES   & $0.44\pm 0.04$ & 0.19 & 1.7  [0.5,2.8] \nl 
 & &  \nl    
Sample 6 (Vis)     & $0.5$    & NO   & $0.45\pm 0.04$ & 0.16 & 1.2 [0.10,2.4] \nl  
Sample 6 (Vis)     & $0.01$   & NO   & $0.43\pm 0.04$ & 0.05 & 2.0 [0.8,3.2] \nl   
Sample 6 (Vis)     & $0.5$    & YES  & $0.43\pm 0.04$ & 0.16 & 2.0 [0.9,3.0] \nl  
Sample 6 (Vis)     & $0.01$   & YES  & $0.40\pm 0.04$ & 0.03 & 2.8 [1.7,4.4] \nl  
 & &  \nl    
Sample 7 (Vis) & $0.5$    & NO   & $0.48\pm 0.04$ & 0.60 & 0.2 [-1.0,1.5] \nl  
Sample 7 (Vis) & $0.01$   & NO   & $0.46\pm 0.04$ & 0.21 & 1.1 [-0.2,2.3] \nl  
Sample 7 (Vis) & $0.5$    & YES  & $0.46\pm 0.04$ & 0.52 & 1.0 [0.0,2.2] \nl  
Sample 7 (Vis) & $0.01$   & YES  & $0.44\pm 0.04$ & 0.24 & 1.9 [0.7,3.1] \nl  

\tablecomments{Sample 1 is composed of all elliptical galaxies (as judged from profile
fitting) with spectroscopic redshifts. Sample 2 adds photometric redshifts to Sample 1.
Sample 3 is Sample 1 plus spectroscopic failures but each failure is assigned a redshift
of $z=0.2$.
Sample 4 is composed of Sample 1 objects but excluding ``possible'' S0 galaxies. Sample
5 adds all S0 galaxies (as judged from profile fitting) to Sample 1. Sample 6 is based solely on
visual classifications from Brinchmann et al. 1998 and Sample 7 adds photometric
redshifts. $P_{ks}$ is the probability, calculated using a Kolmogorov-Smirnov test,
that a model without number density evolution is acceptable. 
The best-fit exponent $\gamma$ [and its $1\sigma$ confidence interval] is given for an
evolutionary law of the form $F \propto (1+z)^{-\gamma}$.
}
\enddata
\end{deluxetable}